\documentclass{article}

\usepackage{amd}
\usepackage{listings}
\usepackage{tcolorbox}
\usepackage{listings}

\usepackage{xcolor}
\usepackage{ifthen} 

\definecolor{defscol}{HTML}{ecd8d7} 
\definecolor{asumscol}{HTML}{ecd8d7} 

\definecolor{rmkscol}{HTML}{313160} 
\definecolor{exmscol}{HTML}{e04b52} 

\definecolor{lemscol}{HTML}{2c3943} 
\definecolor{thmscol}{HTML}{595765} 
\definecolor{prpscol}{HTML}{9c98b1} 
\definecolor{corscol}{HTML}{dfd9fd} 

\definecolor{clmscol}{HTML}{165c58} 
\definecolor{facscol}{HTML}{28a8a1} 

\newtcbtheorem[number within=section]{mydefinition}{Definition}
{
    enhanced,
    frame hidden,
    titlerule=0mm,
    toptitle=1mm,
    bottomtitle=1mm,
    fonttitle=\bfseries\large,
    coltitle=black,
    colbacktitle=defscol!40!white,
    colback=defscol!20!white,
}{defn}

\newtcbtheorem[use counter from=mydefinition]{myassumption}{Code}
{
    enhanced,
    frame hidden,
    titlerule=0mm,
    toptitle=1mm,
    bottomtitle=1mm,
    fonttitle=\bfseries\large,
    coltitle=black,
    colbacktitle=asumscol!40!white,
    colback=asumscol!20!white,
}{asum}

\newcommand{\coolcode}[1]{\texttt{#1}}

\NewDocumentCommand{\asumr}{mm+m}{
    \begin{myassumption}{#1}{#2}
        \coolcode{#3}
    \end{myassumption}
}

\usepackage{microtype}
\usepackage{graphicx}
\usepackage{subfigure}
\usepackage{booktabs} 

\usepackage{placeins}
\usepackage{stfloats}

\usepackage{hyperref}
\usepackage{url}
\usepackage{graphicx}
\usepackage{algorithm}
\usepackage{algorithmic}
\usepackage{multirow}
\usepackage{amsmath}
\usepackage{multirow}  
\usepackage{array}     
\usepackage{amsmath}   
\usepackage{graphicx}  
\usepackage{booktabs}  
\usepackage{amssymb} 
\usepackage{pifont} 
\usepackage{wrapfig}
\usepackage{tikz}
\usepackage{algorithm} 
\usepackage{amsthm}
\usepackage{fvextra}

\usepackage{hyperref}
\usepackage[affil-it]{authblk} 

\usepackage{amsmath}
\usepackage{amssymb}
\usepackage{mathtools}
\usepackage{amsthm}
\usepackage{amsmath}
\usepackage{amssymb}
\usepackage{amsthm}

\usepackage[capitalize,noabbrev]{cleveref}

\theoremstyle{plain}
\newtheorem{theorem}{Theorem}[section]

\newtheorem{lemma}[theorem]{Lemma}

\theoremstyle{definition}
\newtheorem{definition}[theorem]{Definition}

\theoremstyle{remark}

\usepackage[textsize=tiny]{todonotes}
\newenvironment{keywords}
{\noindent \textbf{Keywords:}}
{}

\usepackage{amsmath,amsfonts,bm}

\def\eqref#1{equation~\ref{#1}}

\def\1{\bm{1}}

\DeclareMathAlphabet{\mathsfit}{\encodingdefault}{\sfdefault}{m}{sl}
\SetMathAlphabet{\mathsfit}{bold}{\encodingdefault}{\sfdefault}{bx}{n}

\usepackage{hyperref}
\usepackage{url}

\usepackage{array} 
\title{Chain-Oriented Objective Logic with Neural Network Feedback Control and Cascade Filtering for Dynamic Multi-DSL Regulation}
\author{Jipeng Han}
\affil{OpenImmortal (Beijing) Technology Co., Ltd.\\ \texttt{coolang2022@gmail.com}}

\date{} 
\begin{document}
\maketitle
\begin{abstract}
\textbf{Contributions to AI:} This paper proposes a neuro-symbolic search architecture integrating discrete rule-based logic with lightweight \textbf{Neural Network Feedback Control (NNFC)}. Utilizing \textbf{cascade filtering} to isolate neural mispredictions while dynamically compensating for static heuristic biases, the framework theoretically guarantees search stability and efficiency in massive discrete state spaces. 

\textbf{Contributions to Engineering Applications:} The framework provides a scalable, divide-and-conquer solution coordinating heterogeneous rule-sets in knowledge-intensive industrial systems (e.g., multi-domain relational inference and symbolic derivation), eliminating maintenance bottlenecks and state-space explosion of monolithic reasoning engines.

Modern industrial AI requires dynamic orchestration of modular domain logic, yet reliable cross-domain rule management remains lacking. We address this with \textbf{Chain-Oriented Objective Logic (COOL)}, a high-performance neuro-symbolic framework introducing: (1)~\textbf{Chain-of-Logic (CoL)}, a divide-and-conquer paradigm partitioning complex reasoning into expert-guided, hierarchical sub-DSLs via runtime keywords; and (2)~\textbf{Neural Network Feedback Control (NNFC)}, a self-correcting mechanism using lightweight agents and a \textbf{cascade filtering architecture} to suppress erroneous predictions and ensure industrial-grade reliability. Theoretical analysis establishes complexity bounds and Lyapunov stability.

Ablation studies on relational and symbolic tasks show CoL achieves 100\% accuracy (70\% improvement), reducing tree operations by \textbf{91\%} and accelerating execution by \textbf{95\%}. Under adversarial drift and forgetting, NNFC further improves accuracy and reduces computational cost by 64\%. COOL provides a robust, interpretable, and efficient foundation for modular neuro-symbolic reasoning in industrial engineering\textsuperscript{\textasteriskcentered}.


\end{abstract}

\begin{keywords}
Neuro-symbolic engineering; Domain-specific language regulation; Domain-specific language decomposition; Neural network feedback control; Cascade filtering; Industrial rule-driven control
\end{keywords}

\noindent \textsuperscript{\textasteriskcentered}Project address: \url{https://coolang.org}

\section{Introduction}

\begin{figure}[t!] 
    \centering
    \includegraphics[width=0.85\linewidth]{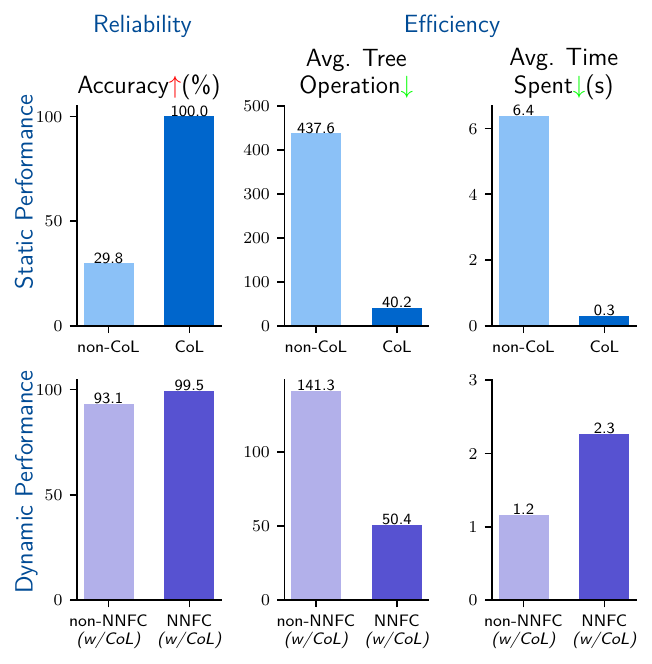}
    \caption{Empirical validation of the neuro-symbolic COOL framework across rule-intensive engineering benchmarks. 
    \textbf{(Top) Static Performance:} The Chain-of-Logic (CoL) achieves structural determinism, elevating accuracy to 100\% while inducing an exponential reduction in state-space complexity (91\% fewer tree operations and 95\% faster latency) compared to the monolithic unregulated baseline. 
    \textbf{(Bottom) Dynamic Performance:} In dynamic (non-stationary) environments, the Neural Network Feedback Control (NNFC) provides critical adaptive compensation. While the distributed neural agents introduce a controlled inference overhead, they guarantee execution-level stability (99\%+ accuracy) where static expert heuristics alone degrade. 
    The results underscore the synergistic advantage of combining deterministic symbolic flows with lightweight, self-correcting AI feedback.}
    \label{fig:performance_overall}
\end{figure}
Domain-Specific Languages (DSLs) are pivotal for encapsulating domain logic in complex engineering fields ranging from robotics to industrial control and automated software synthesis~\cite{dsl_robotic_survey_2014,dsl_robotic_design_2020,dsl_finance_2023}, enabling efficient, verifiable, and rule-intensive system design. The regulation of multiple modular DSLs offers a compelling path toward scalable and reusable system composition. In real-world software engineering, decomposing a monolithic DSL into manageable, task-specific sub-DSLs via a divide-and-conquer strategy is highly encouraged to alleviate tight coupling and maintenance bottlenecks~\cite{dsl_modularization_summary_2005}. This modular paradigm is particularly critical in Industry 4.0 environments, where complex rule-sets must be orchestrated across heterogeneous automated production lines~\cite{industrial_dsl_2022}. However, multi-DSL regulation introduces a fundamental paradox: effective collaboration necessitates the interleaved application of rules from heterogeneous modules, yet this interaction inevitably provokes the fundamental risk of unregulated rule application behaviors (e.g., severe state-space explosion) that undermine system safety, determinism, and reasoning efficiency~\cite{dsl_challengeing_modularization,dsl_modularization_challenge_2013}. This core tension between a necessary mechanism and its detrimental consequences remains unresolved, placing a substantial burden on domain experts who must manually write and maintain intricate, heavily coupled regulation code~\cite{dsl_modularization_summary_2005}.

The root cause of this paradox lies in the misalignment between rule premises and their applicable domains during multi-DSL collaboration. A rule’s premise and specific pattern matching, which are perfectly sufficient within its native, isolated DSL, no longer guarantee correct or efficient application in a broader, heterogeneous context where semantic overlaps between modular components frequently lead to unforeseen logical conflicts~\cite{conflict_detection_2022}. Manual coordination via intermediate code attempts to resolve this by realigning each DSL to the new problem domain. However, this process requires extensive debugging, often necessitates destructive modifications to the original DSLs themselves, and remains highly prone to errors, creating a significant "maintenance debt" that severely restricts the agility of large-scale industrial expert systems~\cite{maintenance_debt_2022}. When introducing AI to automate this regulation, a critical architectural choice emerges. While single, monolithic large neural networks (such as Large Language Models) excel at open-ended, general-purpose tasks, they operate in a fundamentally different paradigm from highly customized, rule-intensive engineering environments where close-ended strategy synthesis and deterministic execution are paramount~\cite{deterministic_ai_2023}. For industrial applications emphasizing low deployment costs, strict execution reliability, and granular modularity, relying on a massive, unified model is often impractical due to its lack of formal transparency and high sensitivity to domain-specific perturbations~\cite{industrial_ai_transparency_2021}. Instead, a "small but many" approach—deploying multiple, lightweight neural networks tightly coupled with specific modular DSLs—provides customized and predictable control without prohibitive overhead and aligns with the current industrial shift toward edge-based modular intelligence~\cite{modular_edge_ai_2024}. Consequently, the problem can be reframed: \textbf{directly and dynamically governing rule application scope through a distributed, lightweight, and self-correcting AI mechanism offers a principled, reliable alternative to manual realignment~\cite{adaptive_neurosymbolic_2022}}.

Our approach to multi-DSL regulation, COOL (Chain-Oriented Objective Logic), intersects with several research areas, yet remains fundamentally distinct from them: Work in \textbf{DSL modularization} focuses on structuring domain logic but lacks mechanisms for dynamic runtime cross-DSL coordination~\cite{dsl_modularization_example_2024, dsl_modularization_exammple_2014}. Research in \textbf{neural adaptive control} (e.g., in robotics or process control) offers learning-based dynamic adaptation but is confined to the continuous state spaces of physical systems, making it inapplicable to the discrete, symbolic reasoning required for coordinating multiple DSLs~\cite{neural_adaptive_control_robot_2025,neural_adaptive_control_reinforcement_2012,neurall_adaptive_feedback_control_2025}. Techniques in \textbf{neural-guided search} effectively compress search spaces but are fundamentally designed for static, monolithic situations and often struggle with the dynamic volatility of multi-domain engineering logic~\cite{dynamic_neural_search_2021}.

\begin{figure}[t!]
    \centering
    \includegraphics[width=0.95\linewidth]{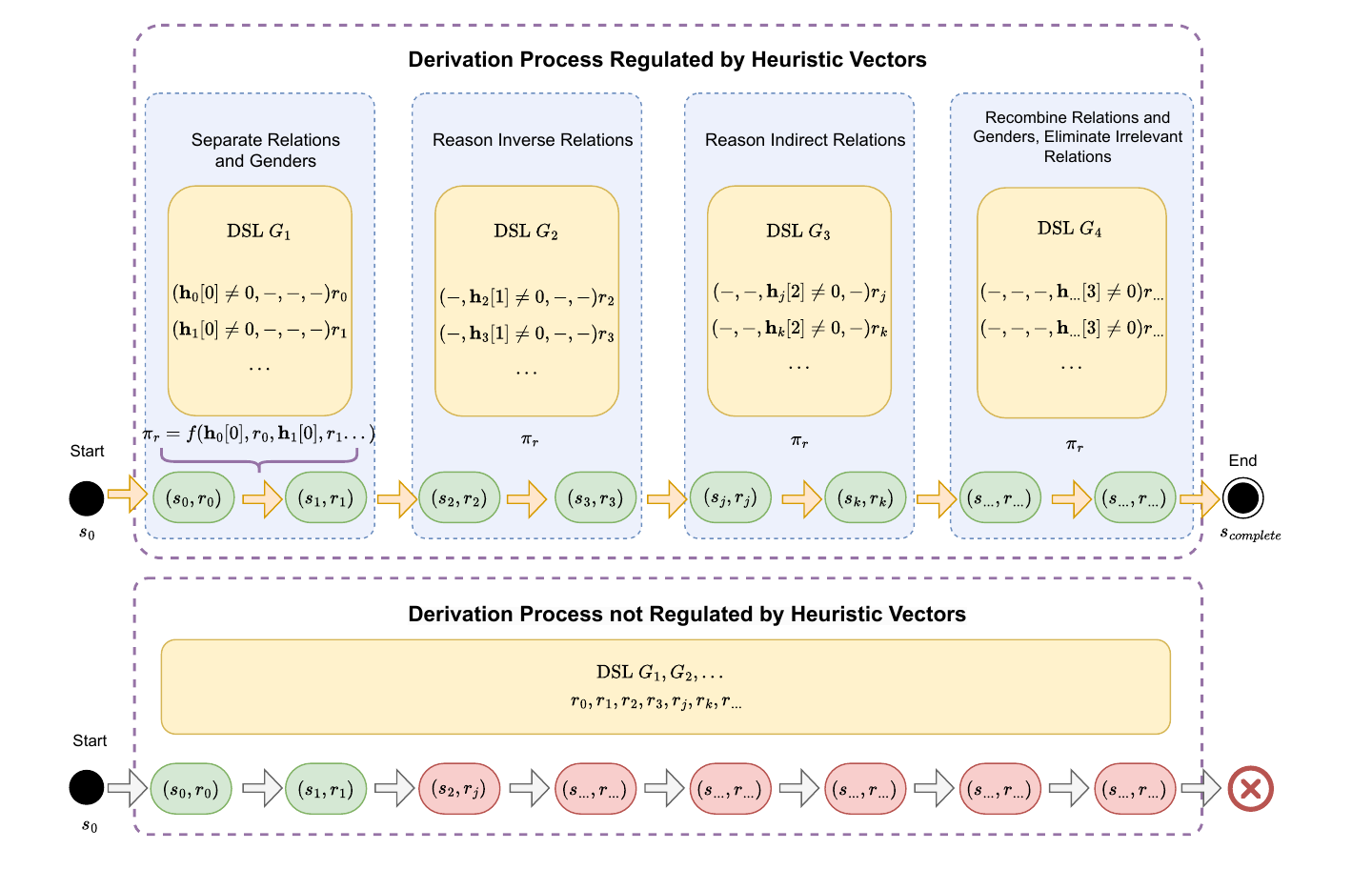}
    \caption{Contrast between regulated and unregulated symbolic derivation processes. \textbf{(Top)} The COOL framework utilizes expert-defined heuristic vectors to partition logic into specialized sub-DSLs ($G_1$ to $G_4$), ensuring a deterministic, stage-wise refinement toward a complete state ($s_{\text{complete}}$). \textbf{(Bottom)} In a monolithic environment without heuristic regulation, the interleaved application of heterogeneous rules leads to unregulated behaviors and search-space explosion, ultimately resulting in derivation failure.}
    \label{fig:heuristic_vector}
\end{figure}

To address these limitations, COOL---the first framework specifically designed for dynamic multi-DSL regulation in complex engineering applications---introduces two novel components that work in concert to precisely and dynamically govern rule application scope and resolve conflicts:

(1) \textbf{Chain-of-Logic (CoL)} dynamically defines the affiliation of rules to specific sub-DSLs. Serving as a divide-and-conquer strategy, CoL allows domain experts to decompose complex reasoning tasks into manageable workflows partitioned into distinct activities, where each logical activity is operationalized by a specialized sub-DSL environment. This structured modularity aligns with the principles of Knowledge-Based Engineering (KBE), where partitioning global domain knowledge into localized, expert-validated activities is essential for maintaining system-wide consistency~\cite{modular_knowledge_2021}. Furthermore, the use of transition operators within this hierarchical structure mirrors the latest methodologies in dynamic task planning for modular industrial systems, ensuring deterministic reachability in complex state spaces~\cite{dynamic_planning_2023}. Its runtime keywords act as transition operators, dynamically switching between or terminating sub-DSLs based on contextual states, thereby preventing conflicting applications across modules without requiring a monolithic design.

(2) \textbf{Neural Network Feedback Control (NNFC)} complements CoL by introducing distributed, lightweight, and modular neural agents. The number of these agents automatically scales with the number of DSLs, providing a flexible intelligence layer that mirrors the scalability requirements of modern cyber-physical production systems~\cite{modular_industrial_control_2023}. They continuously monitor the regulation process and adaptively refine the scope of the rules they oversee. Crucially, NNFC mitigates the instability risks inherent in neural components by employing \textbf{cascade filtering} to detect and block erroneous predictions through discrepancy amplification. This cascade-based verification approach is increasingly recognized as a robust standard for maintaining safety and accuracy in neural-driven industrial controllers~\cite{cascade_verification_2023}. Such multi-stage verification is vital for ensuring the "zero-defect" reliability required in close-ended strategy synthesis for industrial automation~\cite{reliable_industrial_reasoning_2023,eaai_neurosymbolic_2022}. This architecture provides dynamic compensation to the baseline heuristics while guaranteeing execution-level reliability.

By integrating CoL's symbolic precision with NNFC's adaptive robustness, COOL effectively resolves the paradox inherent in multi-DSL regulation, establishing a state-of-the-art paradigm for neuro-symbolic reasoning in high-stakes engineering domains~\cite{neurosymbolic_review_2022}.

To rigorously validate the COOL framework, we integrate theoretical analysis with empirical ablation studies. Theoretically, we establish its foundation through formal methods: for CoL, we prove its expressiveness in encapsulating multi-DSL logic and derive complexity bounds for its regulation mechanism, demonstrating its efficacy in suppressing unregulated behavior emergence and confining the search space via adaptive state-space reduction~\cite{state_space_reduction_2021}; for NNFC, we formalize the feedback control and \textbf{cascade filtering} mechanism, proving its convergence properties and analyzing stability under adaptive conditions using Lyapunov analysis~\cite{lyapunov_neural_control_2021,eaai_stability_2022}. Experimentally, we conduct extensive internal ablation studies within the COOL framework under both static (fixed conditions) and dynamic (evolving conditions) scenarios on rule-intensive engineering tasks. This dual-phase validation follows the established methodology for assessing robust decision-making in complex symbolic environments~\cite{symbolic_validation_2023}. Overall performance (Figure~\ref{fig:performance_overall}) shows that the CoL component alone significantly improves accuracy by 70\%, while reducing tree operations by 91\% and time by 95\% compared to a baseline without CoL's declarative control primitives. In dynamic environments where conditions evolve, NNFC further boosts accuracy by 6\% and cuts tree operations by 64\% compared to the CoL-only variant. Together, these theoretical and empirical results conclusively validate COOL as a highly efficient and reliable framework enabling dynamic multi-DSL regulation through the synergistic combination of CoL's declarative control flow and NNFC's adaptive modulation.

The main contributions are as follows:

\begin{enumerate}

\item \textbf{Chain-of-Logic (CoL):} A structured, divide-and-conquer paradigm for multi-DSL regulation. CoL enables decomposing complex DSLs into manageable sub-DSLs via \textbf{heuristic vectors}, while runtime \textbf{keywords} act as transition operators to govern cross-DSL control flow and suppress unregulated behaviors in large-scale search spaces.

\item \textbf{Neural Network Feedback Control (NNFC):} A self-correcting AI mechanism for rule-intensive engineering. NNFC utilizes distributed, lightweight neural agents to dynamically compensate for expert heuristics. It leverages a \textbf{cascade structure} (sequential network coupling) to amplify prediction discrepancies, filtering erroneous signals to ensure execution stability.

\item \textbf{Theoretical and Empirical Foundation:} We mathematically prove CoL complexity reduction bounds and NNFC Lyapunov stability under dynamic conditions, validating their synergistic integration in the \textbf{COOL} framework through extensive experiments on rule-intensive symbolic and relational tasks.

\end{enumerate}

\begin{figure}[t!]
    \centering
    \includegraphics[width=\linewidth]{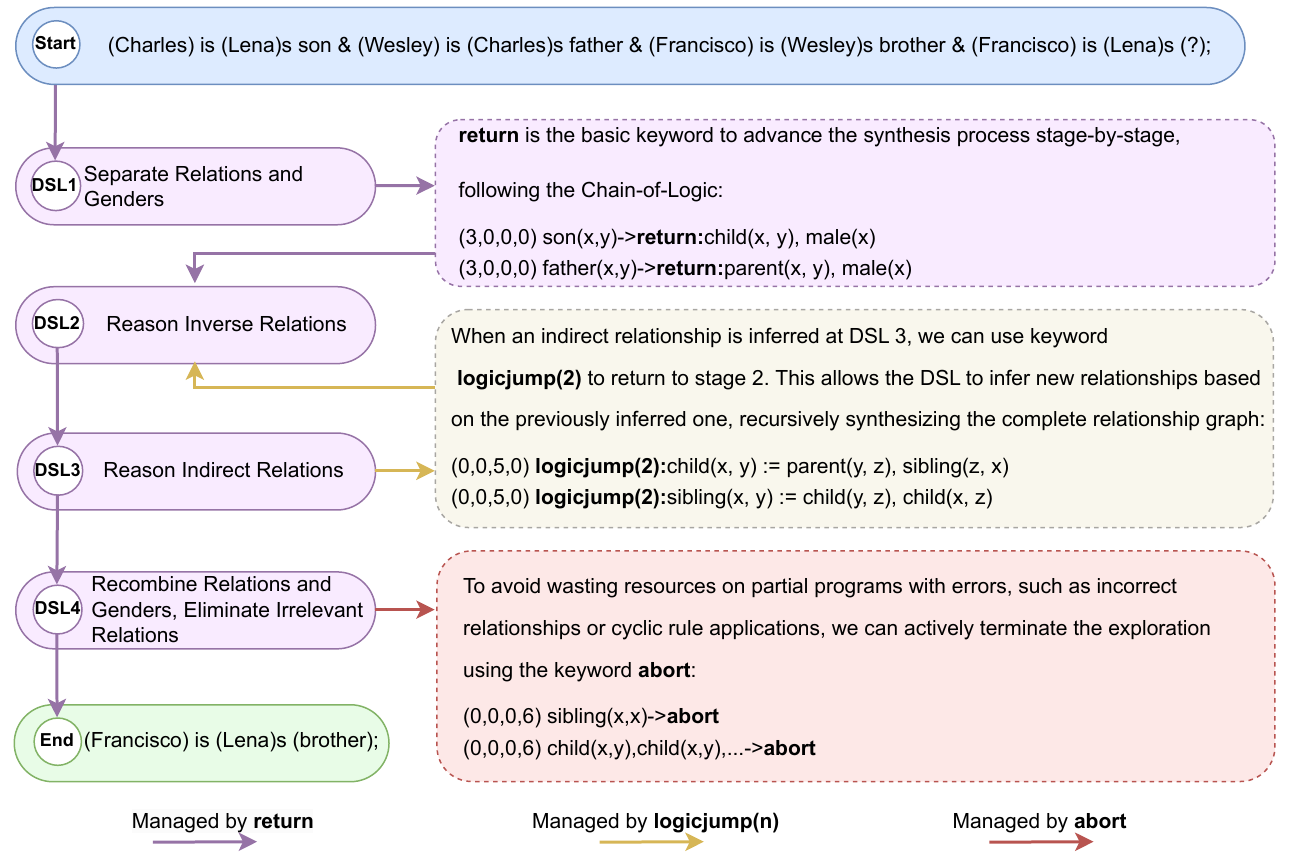}
    \caption{Illustrative keyword-driven DSL transition in Chain-of-Logic. The reasoning process is decomposed into a sequence of expert-defined logical \textbf{activities}, each operationalized by a specialized \textbf{sub-DSL environment}. This architecture enables runtime keywords to precisely govern the transition and dynamic control flow across these modular logic domains.}
    \label{fig:control_words}
\end{figure}

\section{Method}

This section details the implementation of the COOL framework, focusing on the principles of its two core components designed for multi-DSL regulation in rule-intensive engineering environments: the Chain-of-Logic (CoL) and Neural Network Feedback Control (NNFC). Our approach operates on Domain-Specific Languages (DSLs), formally defined as grammatical formalisms. In practical engineering scenarios, such as automated control logic synthesis, the core reasoning process involves the step-wise derivation of symbolic structures from an initial nonterminal through the sequential application of production rules~\cite{gulwani2017program}. This foundational approach utilizes formal grammars to partition complex engineering tasks into well-defined and manageable reasoning stages~\cite{formal_logic_synthesis_2021}. To manage the inherent complexity and severe state-space explosion typical of monolithic DSLs, COOL explicitly adopts a divide-and-conquer strategy: it utilizes CoL for expert-guided structural decomposition and workflow management, aligning with the established principle of decomposition-based complexity reduction in large-scale modular systems~\cite{modular_complexity_2020}, while deploying NNFC to provide lightweight, distributed dynamic compensation during runtime.

\subsection{Chain-of-Logic (CoL)}

The Chain-of-Logic (CoL) serves as an activity-based modular regulation architecture designed to coordinate multiple sub-DSLs that require frequent switching and rule sharing across multi-step engineering tasks. This is achieved through two synergistic mechanisms: \textit{heuristic vectors}, which encode expert knowledge to define rule affiliations, and their runtime keywords (control primitives), which govern deterministic transitions across the DSL workflow.

\paragraph{Heuristic vectors.} As a structured embodiment of domain expertise, a heuristic vector dictates the divide-and-conquer sequence for applying modular logic to a complex task. In COOL, rules are not restricted to a monolithic set; instead, each rule is annotated with a heuristic vector that defines its applicability across different logical activities (operationalized as sub-DSLs $G_1 \dots G_4$ in Figure~\ref{fig:heuristic_vector}). This mechanism facilitates the injection of expert-defined priors into the search process, effectively transforming a high-entropy blind search—which typically leads to the derivation failures shown in the unregulated case—into a structured, knowledge-guided optimization~\cite{expert_heuristics_2017}. The specific non-zero values represent the baseline heuristic tendencies provided by experts. As illustrated in the successful regulated trajectory (Figure~\ref{fig:heuristic_vector}, top), symbolic structures are processed streamingly along this predefined workflow. This architecture allows the system to skip unnecessary downstream stages while preventing unauthorized reversions to upstream ones, thereby effectively restricting the operational search envelope. Theoretically, this constraint minimizes the effective branching factor, suppressing the exponential growth of the state space in deep reasoning chains~\cite{search_space_pruning_2021}.

\paragraph{Keywords (Control primitives).} To enable dynamic and flexible workflow management within the deterministic CoL framework, COOL introduces three essential keywords—\verb|return|, \verb|logicjump(n)|, and \verb|abort|. As exemplified by the multi-stage relational reasoning pipeline in Figure~\ref{fig:control_words}, these primitives function as discrete transition operators that govern the progression, recursion, and pruning of reasoning branches. The operational semantics of these keywords align with the principles of Discrete Event Systems (DES), where transitions are supervised to ensure that the system follows safe and efficient execution pathways~\cite{discrete_event_systems_2021}:

\begin{enumerate}
    \item \textbf{return}: Concludes the current rule application, either maintaining focus within the active sub-DSL or advancing the symbolic state to the subsequent activity in the chain. As shown in Figure~\ref{fig:control_words} (DSL 1), it facilitates the transformation of primitive attributes into structured relations once the local sub-DSL tasks are concluded.
    \item \textbf{logicjump(n)}: Enables non-sequential transitions by routing the symbolic state back to the activity $n$. In the illustrated workflow, this keyword allows for recursive relationship synthesis by jumping from DSL 3 back to DSL 2, effectively constructing deep logical links.
    \item \textbf{abort}: Serves as a deterministic pruning operator.It forcefully terminates branches that exhibit cyclic rule applications or logical inconsistencies (e.g., conflicting relational assignments in DSL 4), thereby preserving computational resources for feasible derivation paths.
\end{enumerate}

\begin{figure}[htbp]
    \centering
    \includegraphics[width=0.95\linewidth]{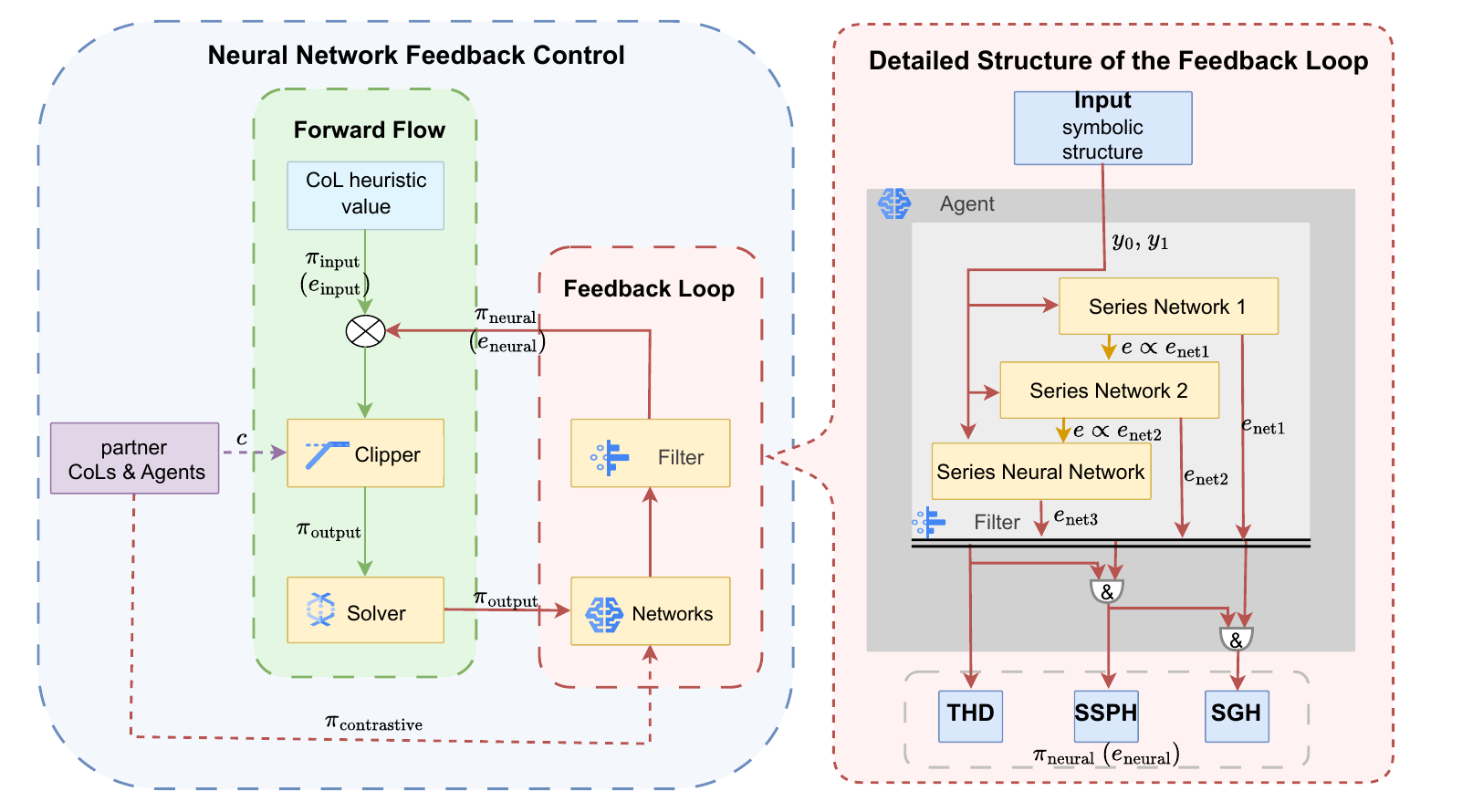}
    \caption{Architecture of the Neural Network Feedback Control (NNFC). \textbf{(Left)} The complete regulation loop: Expert-defined baseline heuristics ($\pi_{\text{input}}$) guide the forward reasoning flow (green path), while distributed neural agents provide dynamic compensation ($\pi_{\text{neural}}$) based on historical reasoning trajectories in the feedback loop (red path). \textbf{(Right)} The cascade filtering mechanism: The neural agent utilizes a sequential network coupling structure to amplify inconsistencies in potential mispredictions. By evaluating output consensus across the \textbf{cascade}, the system filters out high-discrepancy noise and retains only high-quality signals for stable rule regulation. Technical specifications of the lightweight neural architecture, training protocols, and prediction logic are detailed in Appendix~\ref{app:neural}.}
    \label{fig:feedback_with_inner_coupling}
\end{figure}

\subsection{Neural Network Feedback Control (NNFC)}

Neural Network Feedback Control (NNFC) complements the CoL framework by integrating a distributed and lightweight neural compensation layer, enabling continuous adaptation and refinement of the multi-DSL regulation process. Unlike monolithic large-scale models, NNFC is designed for modular engineering environments: its modular neural agents scale with the number of DSLs, ensuring that each task-specific activity receives customized guidance. This architecture adheres to the principle of modular intelligence, where decomposing a global optimizer into localized, low-complexity agents enhances both the scalability and the interpretability of the regulatory policy~\cite{modular_neural_networks_2021}. This architecture supports not only the coordination of homogeneous DSLs within a single task but also the complex management of heterogeneous DSLs across cross-domain engineering scenarios.

NNFC operates by maintaining dedicated, lightweight neural agents for each associated CoL instance. These agents learn from historical reasoning trajectories to provide dynamic compensation to the expert-defined baseline heuristics, effectively refining the rule application scope at runtime. This integration follows the paradigm of informed machine learning, where formal expert priors are utilized to constrain the search space of data-driven models, thereby reducing the risk of stochastic exploration failures~\cite{informed_machine_learning_2021}. As illustrated in Figure~\ref{fig:feedback_with_inner_coupling}, the NNFC architecture is organized into two primary flows:

\textbf{(1) Forward Flow (Guidance and Consensus):} During the reasoning process, the expert-defined CoL signals are superposed with the dynamic compensation signals generated by the neural agents. To guarantee execution-level reliability, a \textbf{Clipper module} (Appendix~\ref{app:clipper}) functions as a consistency guard; it prioritizes control signals that align with the neural-expert consensus while capping or suppressing inconsistent signals that deviate from the established search boundaries. The superposition logic ensures that the neural layer remains an auxiliary compensator rather than a primary driver, preserving the deterministic reachability of the symbolic backbone~\cite{neurosymbolic_determinism_2023}.

\textbf{(2) Feedback Loop (Cascade Filtering and Stability):} The neural agent generates refinement signals based on the current intermediate symbolic constructs. To mitigate the inherent instability risks of neural mispredictions in strictly constrained tasks, NNFC implements a \textbf{cascade filtering mechanism} based on sequential network coupling. This mechanism forces multiple lightweight networks to process the input in series, purposefully \textbf{amplifying discrepancies} in potential erroneous predictions to facilitate deterministic error isolation. Theoretically, this cascade exploits the differences in accumulated computational bias across identical architectures to isolate outliers, effectively acting as a high-pass filter for logical inconsistencies~\cite{cascaded_reliability_2022}. A \textbf{Filter module} then evaluates these discrepancies, blocking low-confidence noise before it can influence the forward flow. 

The architecture of the neural agent is specifically engineered for high-stakes engineering environments where controllability and reliability are paramount. As illustrated in Figure~\ref{fig:feedback_with_inner_coupling} (right), the agent employs multiple cascaded networks with identical architectures. This deliberate design choice simplifies structural analysis and ensures that any induced representation divergence stems exclusively from the serial data flow rather than architectural heterogeneity. By forcing subsequent networks to process both the raw symbolic input and the transformed outputs from their predecessors, the agent actively induces divergent internal representations for the same initial state. This method facilitates a form of cross-network consensus verification, where the stability of a prediction is measured by its invariance~\cite{consensus_learning_2023}. The agent then leverages these filtered, high-confidence signals to provide precise dynamic compensation to the forward reasoning flow.

Each lightweight neural agent functions as a specialized modular plug-in that is automatically loaded alongside its corresponding DSL library. It generates three distinct types of outputs to refine the multi-DSL regulation process. This multi-task output structure allows for the simultaneous evaluation of task-relevance, reachability, and priority, aligning with the multi-objective optimization requirements of automated reasoning systems~\cite{multitask_reasoning_2022}. Figure~\ref{fig:multi_dsl_collaboration} demonstrates how this architecture enhances the regulatory workflow (technical specifications of these output features are summarized in Table~\ref{tab:output_features}):

\begin{figure}[ht]
    \centering
    \includegraphics[width=0.9\linewidth]{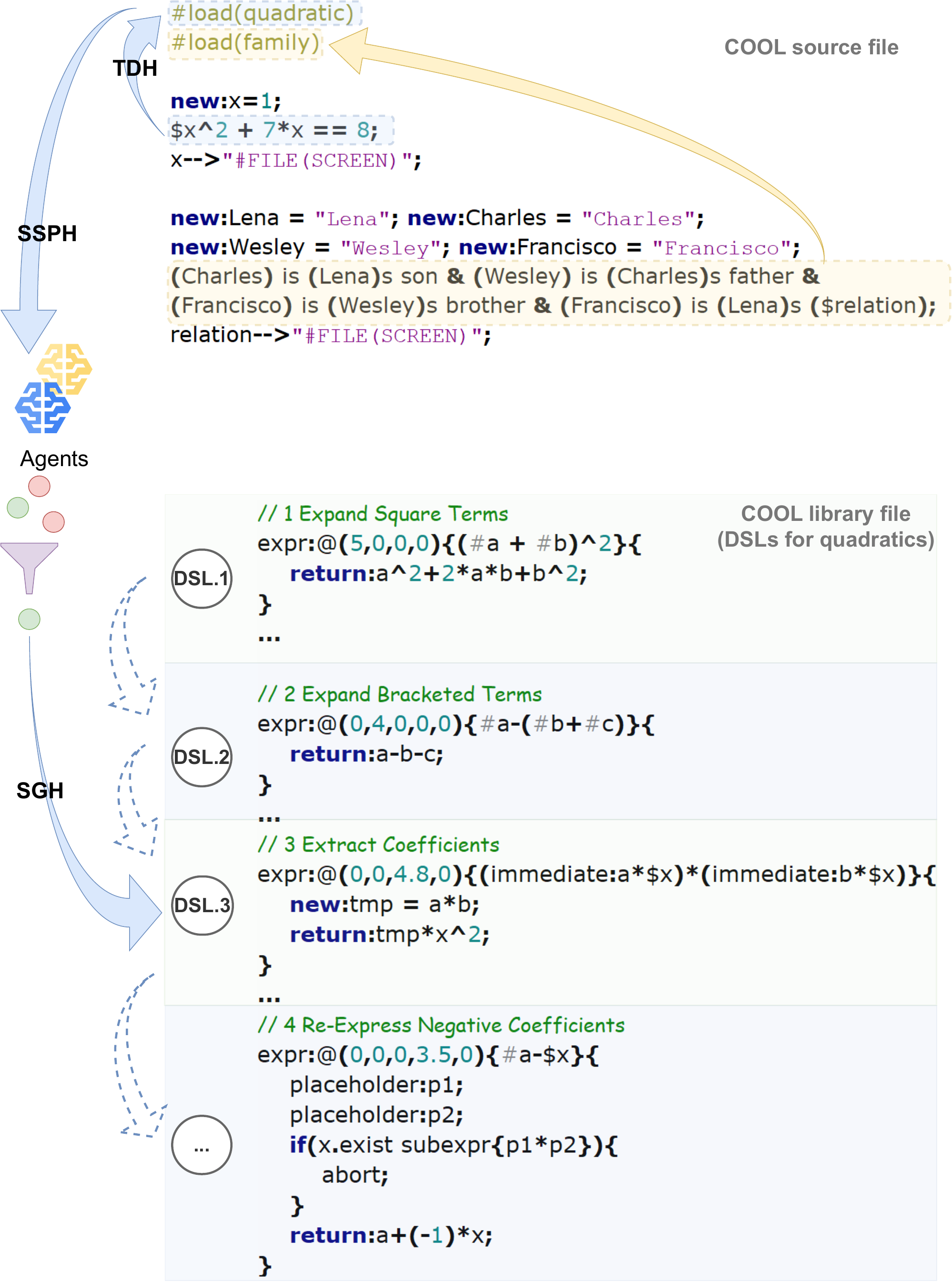}
    \caption{Mechanism of the modular neural agent's multi-head outputs. Upon loading a specific DSL library (e.g., for quadratic symbolic logic), its dedicated neural agent provides localized guidance: (1) \textbf{TDH} identifies the applicable task context to prevent inter-activity interference; (2) \textbf{SSPH} assesses the feasibility of derivation paths to prune the search space; and (3) \textbf{SGH} refines rule priorities to accelerate convergence. This modularity ensures that the AI's influence is strictly confined to relevant tasks, maintaining system-wide determinism.}
    \label{fig:multi_dsl_collaboration}
\end{figure}

\begin{enumerate}
    \item \textbf{Task Detection Head (TDH)}: Functions as a contextual scoping mechanism that determines whether the current symbolic structure falls within the specialized processing domain of the active DSL. This prevents modular interference in cross-task engineering scenarios.
    
    \item \textbf{Search Space Prune Head (SSPH)}: Triggered only when TDH confirms contextual relevance, this head performs a feasibility assessment. It predicts whether a target symbolic structure can be successfully derived from the current state, enabling aggressive search space pruning by eliminating computationally expensive but infeasible branches.
    
    \item \textbf{Search Guidance Head (SGH)}: Active when both context and feasibility are validated, this head provides fine-grained priority compensation. It outputs features of the most promising rules—including the recommended sub-DSL affiliation and optimal heuristic adjustments—to guide the solver toward the most efficient derivation path.
\end{enumerate}
\section{Theoretical Analysis}
\label{sec:theoretical_ana}

We establish the rigorous theoretical foundation of COOL through a formal analysis of its core regulatory mechanisms. This analysis serves two purposes: (1) to define the expressiveness and complexity bounds of the Chain-of-Logic (CoL), ensuring its role as a transparent and efficient orchestrator; and (2) to provide convergence and stability guarantees for the Neural Network Feedback Control (NNFC) using Lyapunov methods. These proofs demonstrate that COOL offers a principled, deterministic solution for governing multi-DSL regulation in safety-critical engineering tasks where maintaining formal consistency across modular boundaries is critical~\cite{modular_consistency_2022}.

\subsection{Expressiveness Integrity of CoL Regulation}

A fundamental requirement for any regulatory framework is that it must not diminish the inherent computational power of the systems it orchestrates. We formally prove that \textbf{CoL preserves the full expressive power of the regulated DSLs}. Since CoL introduces only finite-state coordination mechanisms without requiring additional unbounded auxiliary storage, the composite system's expressiveness is precisely determined by its most expressive constituent DSL. This preservation ensures that the coordination layer does not act as a computational bottleneck~\cite{compositional_expressiveness_2021}. Whether the underlying DSLs are finite-state, context-free, or Turing-complete, CoL acts as a \textit{transparent coordinator} that manages search efficiency without imposing computational constraints on the domain logic itself (see Proof~\ref{app:prf_expressiveness}).

\paragraph{Regulatory Expressiveness Analysis.}
 CoL's regulatory expressiveness—defined by the class of regulatory strategies it can enact—adaptively scales in sync with the computational hierarchy of the governed DSLs. Rather than being a rigid sequence, the control flow paradigm functions as a closed-loop regulator. It enables sophisticated, non-linear regulation strategies that suppress unauthorized rule applications and optimize execution paths by dynamically parameterizing control keywords based on real-time DSL operational states. Such context-dependent regulatory scaling is essential for orchestrating multi-tiered formal languages within autonomous systems~\cite{adaptive_formal_systems_2023}. Specifically, CoL's regulatory expressiveness is not predefined; it scales correspondingly through finite-state, context-free, and Turing-complete levels depending on the feedback capabilities of its constituent DSLs. This ensures that COOL can coordinate tasks across varying computational hierarchies without imposing structural constraints on the underlying logic (see Proof~\ref{app:prf_regulatory_scaling}).

\paragraph{Complexity Analysis.} 
To ensure computational tractability in engineering, CoL systematically mitigates the state-space explosion through three algorithm-agnostic mechanisms governed by control primitives. For a global rule set of size $R$ and an unguided path length $L_{\text{global}}$, the baseline search complexity is $O(R^{L_{\text{global}}})$, which is typically prohibitive for real-time applications due to the NP-hard nature of unrestricted symbolic reasoning~\cite{synthesis_complexity_2021}. CoL transforms this into a regulated search problem, achieving exponential reduction through: (1) \texttt{return} and \texttt{logicjump(n)}, which dynamically restrict the active search envelope, thereby reducing the effective branching factor $b_t$ at each step $t$; and (2) \texttt{abort}, which performs deterministic branch elimination with a multiplicative pruning factor $(1-p_t)$.

The resulting composite complexity is formulated as $O\left( \prod_{t=1}^{L} \left[ (1 - p_t) \cdot b_t \right] \right)$ for a regulated path length $L$. As derived in Proof~\ref{app:prf_complexity}, this complexity is strictly bounded between the unconstrained worst-case $O(R^L)$ and the theoretical linear-time limit $\Omega(L)$ (perfect regulation). The transition from exponential to near-linear complexity demonstrates the effectiveness of structural decomposition in high-dimensional discrete search environments~\cite{search_tractability_2023}. In practice, the exponential restriction of the branching factor and the multiplicative pruning effect compound to suppress unregulated search expansion, with the overall efficiency being a direct function of the precision of expert-defined heuristic vectors and runtime keywords.
\subsection{Neural Network Feedback Control (NNFC) Mechanism}
\label{sec:nnfc_theory}
\textbf{Hybrid Heuristic Superposition:} To ensure a reliable integration of AI and expert logic, NNFC treats neural guidance as a dynamic compensation term added to a deterministic backbone. Let the fixed expert-defined heuristics be $\mathbf{\pi}_{\text{input}}$ and the neural agent's output be $\mathbf{\pi}_{\text{neural}}$. A hybrid strategy $\mathbf{\pi}_{\text{output}}$ is generated through a superposition algorithm:
\begin{equation}
    \mathbf{\pi}_{\text{output}} = \mathbf{\pi}_{\text{input}} + \lambda \cdot \mathbf{\pi}_{\text{neural}}
\end{equation}
where $\lambda$ is a mixing coefficient representing the system's reliance on neural guidance. This additive structure follows the established paradigm of knowledge-based compensation, where learned neural signals serve as local residuals to refine static expert priors~\cite{informed_ml_2021}. To maintain long-term adaptability, the lightweight neural agents are continuously refined via gradient descent on historical reasoning trajectories.

\textbf{Composability and Safety Guarantees:} We formally establish that neural oracle guidance can be safely composited with symbolic heuristics without compromising algorithmic completeness. For engineering systems requiring strict optimality, the superposition preserves admissibility as long as the neural agent aligns with the true cost boundaries. Crucially, under the "small but many" modular architecture, even in the presence of erroneous neural predictions, the symbolic backbone ensures that the search space remains reachable, consistent with the principles of reachability analysis in modular formal synthesis~\cite{gulwani2017program}, thereby guaranteeing that a valid solution will eventually be found—albeit with a potential temporary impact on search efficiency (see Proof~\ref{app:prf_composability}). In resource-constrained engineering environments, this compositing effectively increases the probability of successful derivation and minimizes expected execution time.

\paragraph{Cascade Error Amplification and Filtering.} 
To enhance the reliability of the neural agents, NNFC employs a cascade structure that intentionally amplifies internal inconsistencies via error propagation. This series coupling design functions as a discrepancy amplifier, utilizing serial error accumulation as a structural surrogate for traditional fault detection and isolation in high-stakes automation~\cite{fault_detection_2006}. This approach significantly increases the sensitivity of error detection compared to traditional parallel ensembles, especially in low-error regimes. The system defines an inconsistency metric $D$ based on the divergent representations across the cascade. We formally prove that the filtering probability follows an adaptive scaling law: $p_{\text{filter}} \propto \beta \cdot \epsilon \cdot \frac{1 - \gamma^K}{1 - \gamma}$ , ensuring that the regulatory mechanism automatically strengthens its filtering sensitivity as the neural error rate increases, thereby protecting the symbolic CoL workflow from stochastic fluctuations (see Proof~\ref{app:prf_cascade_filtering}).

\paragraph{System Stability and Convergence.} 
The stability of the NNFC framework is governed by a synergistic mechanism of dynamic error control and cascade filtering. We formally prove that the system converges to a bounded stable region. By defining a composite Lyapunov energy function $V = V_{\text{neural}} + V_{\text{output}}$, we analyze the energy dissipation $\Delta V$ under the influence of training efficiency, dynamic drift ($e_{\text{drift}}$), and catastrophic forgetting ($e_{\text{forget}}$). Our analysis yields a critical stability condition: $e_{\text{dynamic}} \leq \frac{\alpha \kappa}{4\beta}$ , where $\alpha$, $\beta$, and $\kappa$ represent coefficients of learning, error injection, and termination sensitivity, respectively. Crucially, the \textbf{cascade filtering} structure safeguards system stability by maintaining a minimum activity termination rate $\rho_{\text{min}} > 0$. This provides a "reliability safety net" that preserves learning opportunities and prevents system divergence even during high-uncertainty neural phases (see Proof~\ref{app:prf_lyapunov} for a detailed Lyapunov stability analysis~\cite{neural_control_system_stability_analysis_method}).

\section{Experiments}

We evaluate the COOL framework through a two-phase empirical validation: (1)~\textbf{Static experiments} under fixed, deterministic conditions to isolate the regulatory efficacy of the Chain-of-Logic (CoL); and (2)~\textbf{Dynamic experiments} involving evolving problem domains and noisy environments to assess the adaptive compensation capabilities of the Neural Network Feedback Control (NNFC). This structured evaluation directly mirrors the theoretical stability and complexity bounds established in Section 3.

\subsection{Experimental Setup}

\textbf{Execution Environment.} To achieve native support for heuristic vector management and keyword-driven control flow with minimal runtime overhead, we implemented a dedicated \textbf{COOL host language}. This environment provides high-performance symbolic manipulation and seamless integration with distributed neural agents. All experiments were executed on a workstation equipped with an Intel i7-14700 CPU, a GTX 4070 GPU (for neural inference), and 48GB of RAM. Detailed language specifications and the design rationale for its low-latency execution engine are provided in Appendix~\ref{app:cool_language}.

\begin{table}[ht]
\centering
\caption{Configuration of rule-intensive engineering benchmarks. Relational tasks are categorized by logical inference depth (edges), while symbolic tasks are defined by expression tree complexity (nodes).}
\setlength{\tabcolsep}{4pt} 
\renewcommand{\arraystretch}{1.1} 
\begin{tabular}{m{0.2\textwidth}m{0.35\textwidth}m{0.35\textwidth}}
\toprule 
\centering \textbf{Task Category} & \centering \textbf{Difficulty Level A} & \centering \textbf{Difficulty Level B} \tabularnewline
\midrule 

\centering Relational Inference & \centering 300 tasks (3-edge logic depth) & \centering 200 tasks (4-edge logic depth) \tabularnewline

\centering Symbolic Transformation & \centering 300 tasks ($\approx$5-node complexity) & \centering 200 tasks ($\approx$9-node complexity) \tabularnewline
\bottomrule
\end{tabular}
\label{tab:tasks_settings}
\end{table}
\textbf{Benchmarks.} We selected two distinct categories of rule-intensive engineering tasks (summarized in Table~\ref{tab:tasks_settings} and exemplified in Appendix~\ref{app:experiment}):
\begin{itemize}
    \item \textit{Relational Reasoning:} These tasks require multi-step inference over heterogeneous knowledge domains, testing the framework's ability to switch between sub-DSLs representing different logical relations.
    \item \textit{Symbolic Computation:} These tasks involve the automated synthesis of calculation steps for non-standard quadratic equations using numerous mathematical transformation laws, testing the framework's precision in high-depth search trees.
\end{itemize}

\textbf{Evaluation Metrics.} To provide a comprehensive assessment of the neuro-symbolic regulation efficiency, we employ the following metrics:
\begin{enumerate}
    \item \textbf{Accuracy (\%):} The primary reliability indicator, measuring the proportion of tasks correctly resolved within resource limits (defined as a maximum of 1,000 rule applications and a maximum path length of 50).
    \item \textbf{Tree Operations (Ops):} A measure of search efficiency, representing the total number of production rules applied.
    \item \textbf{Memory Overhead (Pairs):} Quantified by the number of transformation pairs (symbolic structure $\times$ rule) stored during the search process.
    \item \textbf{GPU Overhead (Inferences):} The total count of neural network invocations, reflecting the computational cost of the "lightweight AI" layer.
    \item \textbf{Time Overhead (s):} The actual wall-clock time required for multi-DSL regulation and task completion.
\end{enumerate}

\paragraph{Experimental Grouping and Ablation Design.} 
Our empirical evaluation is structured around a multi-dimensional ablation study designed to isolate the performance gains of each neuro-symbolic component. The study evaluates two primary dimensions of the proposed framework: (1)~\textbf{Structural Determinism}, verified by comparing regulated (CoL) versus unregulated (Direct) reasoning processes in static environments; and (2)~\textbf{Adaptive Reliability}, assessed by measuring the impact of Neural Network Feedback Control (NNFC) and \textbf{cascade filtering} in dynamic, noisy scenarios. Expert-defined heuristic vectors and the \textbf{neural reliability layer} serve as secondary variables to delineate the contribution of domain knowledge versus AI-driven compensation.

The comprehensive configuration for all experimental groups is detailed in Table~\ref{tab:group_configurations}.

\begin{table}[htbp]
\centering
\caption{Detailed group configurations for the ablation study. Groups marked with \ding{72} represent the full COOL framework implementation, while those with \ding{73} are designated for component-wise ablation. \textbf{Direct} refers to the monolithic unregulated baseline; \textbf{Flt} denotes the activation of the cascade filtering structure.}
\setlength{\tabcolsep}{4pt}
\begin{tabular}{m{0.30\textwidth}|m{0.20\textwidth}m{0.15\textwidth}m{0.10\textwidth}m{0.14\textwidth}}
\toprule
\centering \textbf{Regulatory Group} & \centering \textbf{Environment} & \centering \textbf{Neural Prior} & \centering \textbf{NNFC} & \centering \textbf{Cascade Filtering} \tabularnewline
\midrule 

\centering Direct & \centering static & \centering & \centering & \centering \tabularnewline

\centering \ding{73}Direct (Heuristic) & \centering static & \centering & \centering & \centering \tabularnewline

\centering \ding{72}CoL  & \centering static, dynamic & \centering & \centering & \centering \tabularnewline

\centering \ding{73}Direct + NN & \centering static & \centering \checkmark & \centering & \centering \tabularnewline

\centering \ding{73}Direct (Heuristic) + NN & \centering static & \centering \checkmark & \centering & \centering \tabularnewline

\centering \ding{73}CoL + NN & \centering static & \centering \checkmark & \centering & \centering \tabularnewline

\centering \ding{73}CoL + NNFC & \centering dynamic & \centering & \centering \checkmark & \centering \tabularnewline

\centering \ding{73}Direct + NN (Flt) & \centering static & \centering \checkmark & \centering & \centering \checkmark \tabularnewline

\centering \ding{73}Direct (Heuristic) + NN (Flt) & \centering static & \centering \checkmark & \centering & \centering \checkmark \tabularnewline

\centering \ding{73}CoL + NN (Flt) & \centering static & \centering \checkmark & \centering & \centering \checkmark \tabularnewline

\centering \ding{72}CoL + NNFC (Flt) & \centering dynamic & \centering & \centering \checkmark & \centering \checkmark \tabularnewline
\bottomrule
\end{tabular}
\label{tab:group_configurations}
\end{table}

\subsection{Static Experiments: Evaluating Structural Determinism}

To isolate the impact of the Chain-of-Logic (CoL) on search efficiency, we first conduct static experiments under deterministic conditions (fixed domains and difficulty levels). These experiments aim to verify whether the structural regulation provided by CoL can transform computationally intractable "monolithic" search problems into manageable "modular" tasks.

\begin{table}[htbp]
\centering
\caption{Static performance comparison between the Monolithic Unregulated Baseline (Direct) and the proposed Chain-of-Logic (CoL) across relational and symbolic tasks. CoL achieves deterministic reliability (100\% accuracy) while drastically reducing computational overhead.}
\label{tab:table_for_static_experiment}
\setlength{\tabcolsep}{4pt}
\renewcommand{\arraystretch}{1.1}
\begin{tabular}{m{0.12\textwidth}|m{0.15\textwidth}|m{0.15\textwidth}m{0.15\textwidth}m{0.15\textwidth}m{0.15\textwidth}}
\toprule 
\centering \textbf{Task Type} & \centering \textbf{Regulatory Group} & \centering \textbf{Accuracy \tikz{\draw[red, -latex] (0,0) -- (0,0.3);} (\%)} & \centering \textbf{Avg. Tree Ops \tikz{\draw[green, -latex] (0,0.3) -- (0,0);} } & \centering \textbf{Avg. Memory \tikz{\draw[green, -latex] (0,0.3) -- (0,0);} } & \centering \textbf{Avg. Latency \tikz{\draw[green, -latex] (0,0.3) -- (0,0);} (s)} \tabularnewline
\midrule 

\multirow{2}{*}{\centering Relational} & \centering Direct & \centering 11.3 & \centering 463.9 & \centering 1432.2 & \centering 9.43 \tabularnewline
 & \centering CoL & \centering \textbf{100.0} & \centering \textbf{46.6} & \centering \textbf{177.8} & \centering \textbf{0.48} \tabularnewline
\midrule

\multirow{2}{*}{\centering Symbolic} & \centering Direct & \centering 48.3 & \centering 411.2 & \centering 2285.3 & \centering 3.31 \tabularnewline
 & \centering CoL & \centering \textbf{100.0} & \centering \textbf{33.8} & \centering \textbf{92.7} & \centering \textbf{0.11} \tabularnewline
\bottomrule
\end{tabular}
\end{table}

The empirical results summarized in Table~\ref{tab:table_for_static_experiment} demonstrate that \textbf{CoL provides a deterministic guarantee for multi-DSL coordination while minimizing system overhead}. In both relational and symbolic benchmarks, CoL elevates the task success rate from a baseline below 50\% to a perfect 100\%. More significantly, the structural regulation induces a drastic reduction in computational cost: for relational tasks, tree operations and execution latency are reduced by 90\% and 95\%, respectively; for symbolic tasks, the reduction is even more pronounced, with a 92\% drop in rule applications and a 97\% decrease in time spent. These findings \textbf{empirically validate the complexity reduction bounds derived in our theoretical analysis (Section~\ref{sec:theoretical_ana})}, confirming that restricting the search space through sub-DSLs effectively neutralizes the state-space explosion~\cite{state_space_mitigation_2023}.

\begin{figure}[htbp]
    \centering
    \includegraphics[width=\linewidth]{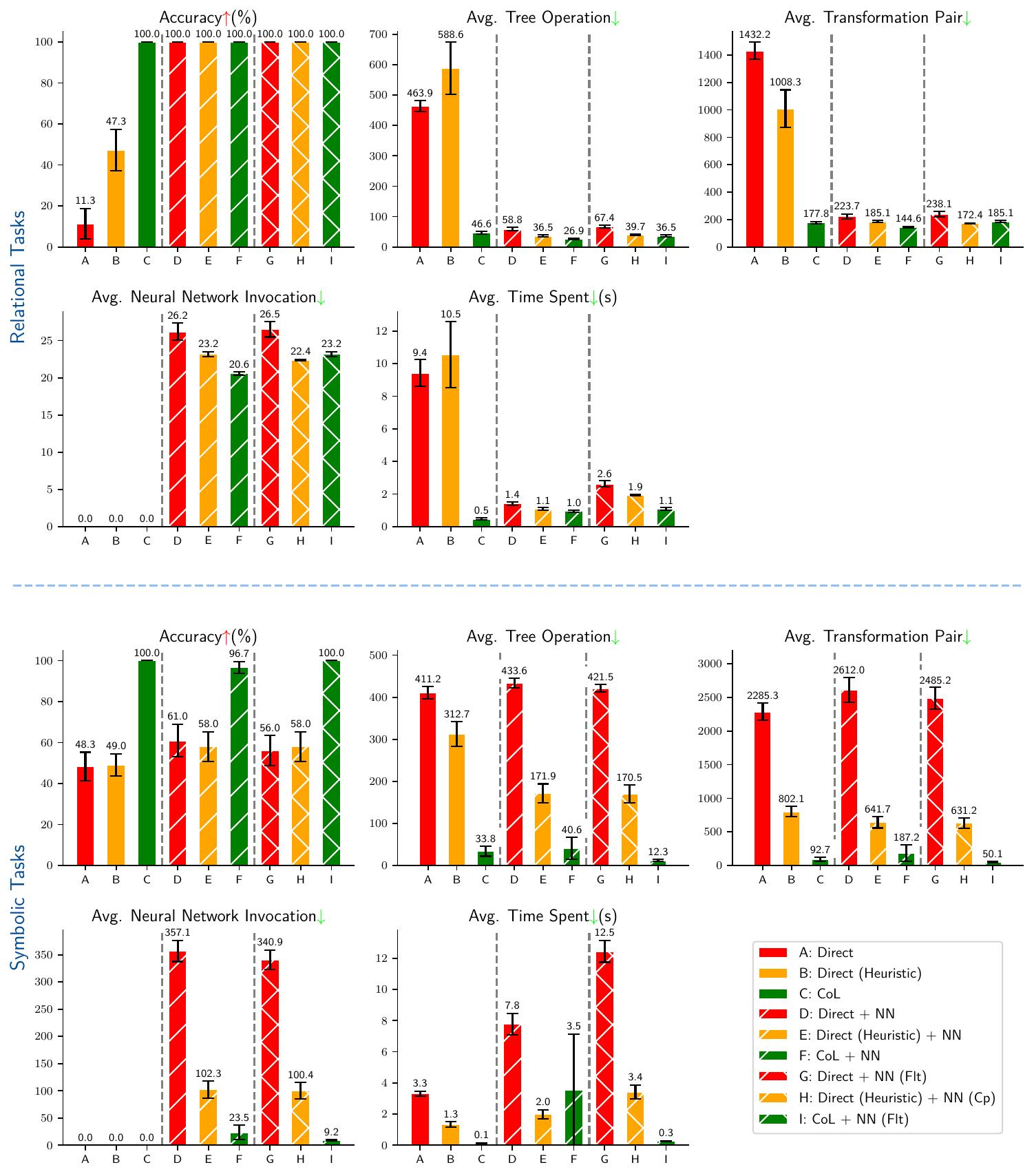}
    \caption{Comparative static performance analysis on relational and symbolic benchmarks. The error bars represent 95\% confidence intervals across 6 independent batches. Results highlight the robust advantage of CoL-based regulation over traditional monolithic approaches.}
    \label{fig:static_combine}
\end{figure}

Further secondary ablation studies (illustrated in Figure~\ref{fig:static_combine}) provide deeper insights into the framework's components:

\textbf{1. Superiority of Divide-and-Conquer over Global Heuristics:} While global heuristics (\textit{Direct (Heuristic)}) improve performance over a random baseline, the \textbf{Heuristic Vector} mechanism in CoL significantly surpasses them in all metrics. This confirms that \textit{explicit activity decomposition}—partitioning the synthesis process into discrete, expert-defined activities—yields far greater performance gains than unorganized heuristic guidance~\cite{modular_decomposition_2022}.

\textbf{2. Theoretical Composability of AI and Symbolic Logic:} The integration of CoL with lightweight neural agents (\textit{CoL + NN}) further narrows the search space, reducing tree operations by an additional 43\% in relational tasks and 64\% in symbolic tasks. This validates the \textbf{composability} proved in Section~\ref{sec:nnfc_theory}, showing that neural guidance effectively compensates for expert-defined baseline heuristics without compromising the search's symbolic integrity~\cite{neurosymbolic_search_2021}.

\textbf{3. Context-Dependent Utility of Filtering:} The cascade filtering group exhibits varied utility based on the task's error tolerance. In \textit{Symbolic Transformation}, where the state space is rigid and error tolerance is near zero, the filtering layer is indispensable for maintaining high accuracy. Conversely, in \textit{Relational Inference} where the task allows for higher error tolerance, the \textbf{stringent filtering mechanism} might occasionally incur a minor penalty by discarding valid but low-confidence outcomes. This highlights the importance of \textbf{cascade filtering} as a tunable reliability guard for safety-critical applications~\cite{cascade_verification_2023}.

\begin{table}[htbp]
\centering
\caption{System performance under non-stationary dynamic conditions. NNFC provides critical adaptive compensation, maintaining near-perfect reliability even as the static baseline (CoL DSL) degrades due to environmental drift.}
\label{tab:table_for_dynamic_experiment}
\setlength{\tabcolsep}{2pt}
\renewcommand{\arraystretch}{1.1}
\begin{tabular}{m{0.13\textwidth}|m{0.18\textwidth}|m{0.11\textwidth}m{0.11\textwidth}m{0.13\textwidth}m{0.13\textwidth}m{0.12\textwidth}}
\toprule
\centering \textbf{Task Domain} & \centering \textbf{Regulatory Group} & \centering \textbf{Accuracy \tikz{\draw[red, -latex] (0,0) -- (0,0.3);} (\%)} & \centering \textbf{Avg. Tree Ops \tikz{\draw[green, -latex] (0,0.3) -- (0,0);} } & \centering \textbf{Avg. Memory \tikz{\draw[green, -latex] (0,0.3) -- (0,0);} } & \centering \textbf{GPU Inferences \tikz{\draw[green, -latex] (0,0.3) -- (0,0);} } & \centering \textbf{Avg. Latency \tikz{\draw[green, -latex] (0,0.3) -- (0,0);} (s)} \tabularnewline
\midrule

\multirow{2}{*}{\centering Relational} & \centering CoL & \centering \textbf{100.0} & \centering 70.0 & \centering 259.8 & \centering - & \centering \textbf{1.05} \tabularnewline
 & \centering CoL+NNFC (Flt) & \centering \textbf{100.0} & \centering \textbf{54.6} & \centering \textbf{224.5} & \centering 21.7 & \centering 2.08 \tabularnewline
\midrule

\multirow{2}{*}{\centering Symbolic} & \centering CoL & \centering 82.6 & \centering 233.5 & \centering 977.1 & \centering - & \centering 1.42 \tabularnewline
 & \centering CoL+NNFC (Flt) & \centering \textbf{99.4} & \centering \textbf{50.3} & \centering \textbf{222.2} & \centering 21.6 & \centering \textbf{1.12} \tabularnewline
\midrule

\multirow{2}{*}{\parbox{1.5cm}{\centering multi-\\domain}} & \centering CoL & \centering 97.5 & \centering 115.2 & \centering 367.6 & \centering - & \centering \textbf{0.99} \tabularnewline
 & \centering CoL+NNFC (Flt) & \centering \textbf{99.0} & \centering \textbf{45.6} & \centering \textbf{250.5} & \centering 72.8 & \centering 3.91 \tabularnewline
\bottomrule
\end{tabular}
\end{table}

\subsection{Dynamic Experiments: Resilience in Non-stationary Environments}

In real-world engineering, regulatory systems must operate in non-stationary environments where the underlying task distribution or system state may drift over time~\cite{concept_drift_review_2020}. We evaluate the Neural Network Feedback Control (NNFC) under varying domains and evolving neural priors to assess its adaptive compensation capabilities.

The results in Table~\ref{tab:table_for_dynamic_experiment} confirm that \textbf{NNFC successfully compensates for the limitations of static expert heuristics in dynamic scenarios, thereby fulfilling the stability guarantees established in our Lyapunov analysis}. As task complexity increases and cross-domain heterogeneous regulation scenarios emerge, the accuracy of the static CoL baseline begins to decline. In contrast, the NNFC-enhanced framework maintains a robust success rate of over 99\%, demonstrating superior adaptability~\cite{dynamic_stability_2022}. Moreover, NNFC significantly optimizes search efficiency, reducing tree operations by 22\% and memory-intensive state transitions by 14\% compared to the CoL-only group. Notably, in symbolic computation tasks, despite the added latency of neural inference, NNFC reduces the overall execution time by 21\%, proving that the AI-driven search space compression outweighs the neural computational cost~\cite{search_compression_2022}.

\begin{figure}[htbp]
    \centering
    \includegraphics[width=0.95\linewidth]{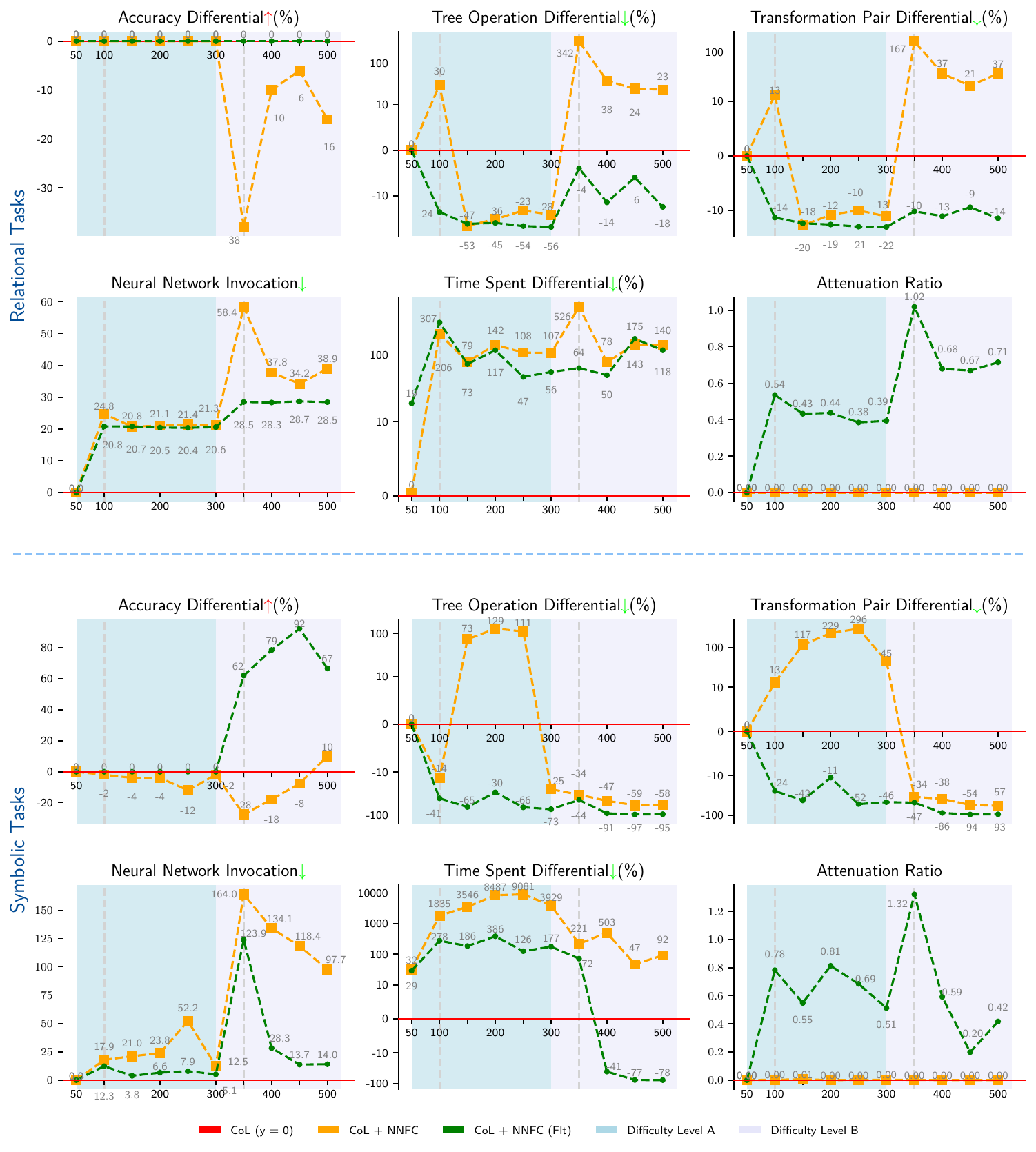}
    \caption{Dynamic performance differential under single-domain evolution. While the unfiltered NNFC suffers from accuracy drops during neural retraining batches, the cascade filtering structure maintains a smooth reliability curve, shielding the symbolic process from training-phase volatility.}
    \label{fig:dynamic_combine}
\end{figure}

\begin{figure}[htbp]
    \centering
    \includegraphics[width=0.95\linewidth]{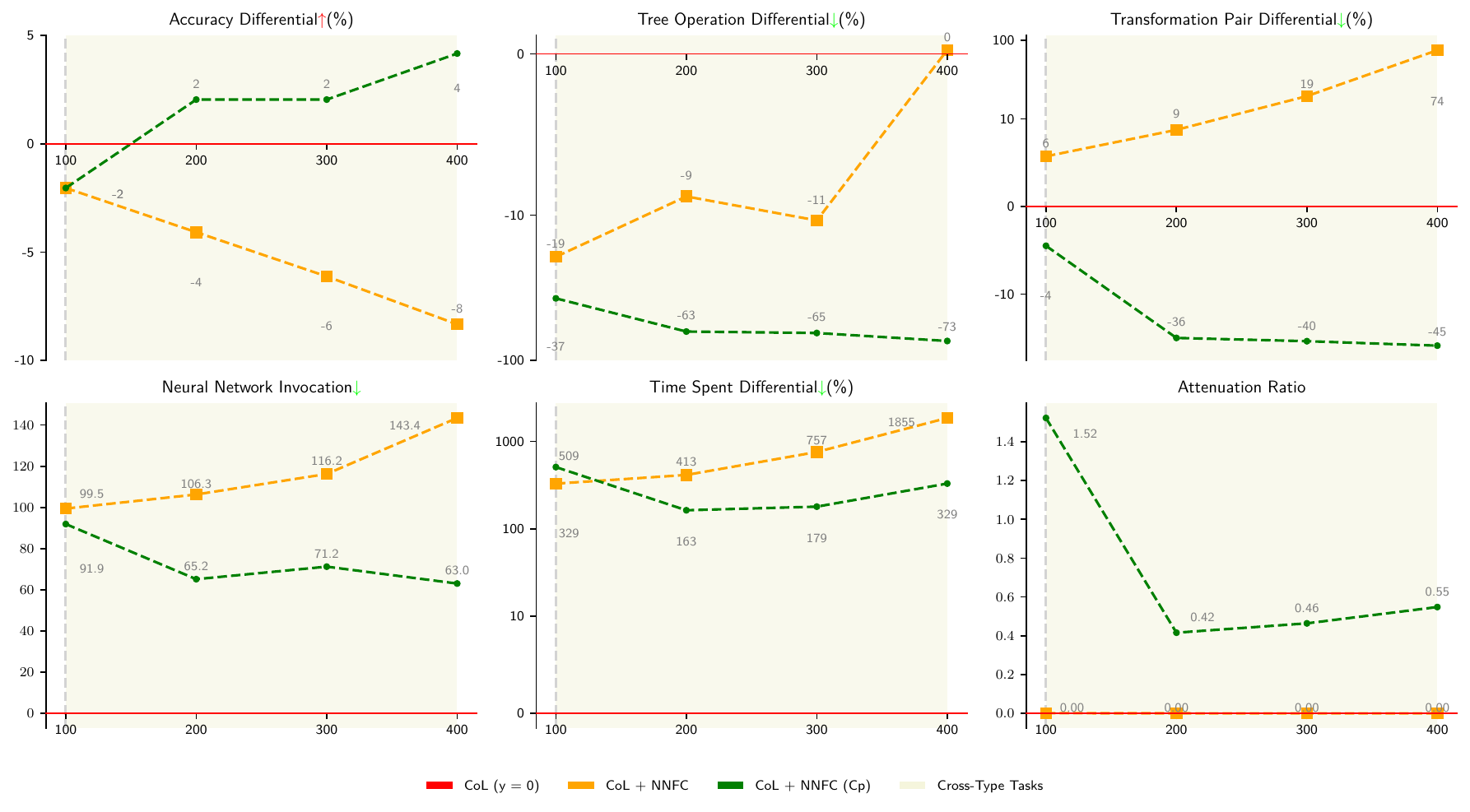}
    \caption{Resilience under cross-domain heterogeneity and catastrophic forgetting. The full COOL framework (CoL + NNFC + Filter) effectively manages the transition between relational and symbolic domains, quickly recovering from the accuracy dips that plague unregulated adaptive systems.}
    \label{fig:dynamic_multidomain}
\end{figure}

Our multi-phase dynamic evaluation (Figures~\ref{fig:dynamic_combine} and \ref{fig:dynamic_multidomain}) further underscores the following engineering insights:

\textbf{1. Necessity of Cascade Filtering for System Stability:} The experimental data reveals that \textit{filtering is indispensable for maintaining search integrity}. As shown in the "attenuation ratio" metrics, the cascade filtering structure successfully blocks over 94\% of misprediction-induced failures. In groups where the filter was disabled, even a single neural error could propagate through the derivation chain, leading to catastrophic accuracy collapses.

\textbf{2. Robustness Against Neural Failure Modes:} We specifically tested COOL under three adversarial neural regimes: (i)~\textbf{Inadequate Learning} (Figure~\ref{fig:dynamic_combine}, Tasks 51-100), where data is sparse~\cite{sparse_learning_2021}; (ii)~\textbf{Generalization Gaps} (Tasks 301-350), involving extreme task difficulty; and (iii)~\textbf{Catastrophic Forgetting} (Figure~\ref{fig:dynamic_multidomain}, Tasks 1-100), where new domain knowledge erodes historical priors~\cite{continual_learning_industrial_2023}. In all cases, when the neural agent's accuracy drops, the cascade structure detects the rising inconsistency and the "attenuation ratio" spikes. This ensures that the system temporarily reverts to the safe CoL baseline, preventing the "hallucination" of incorrect derivation paths. 

\textbf{3. Adaptive Influencer Dynamics:} As the neural agent stabilizes and accuracy improves (e.g., Figure~\ref{fig:dynamic_combine}, Tasks 101-300), the filter's attenuation ratio naturally decreases. This transition marks the point where the AI moves from a "potential liability" to an "efficiency driver," providing precise compensation that accelerates convergence. This adaptive process validates our theoretical conclusion that filtering acts as a critical safety net, allowing the system to harness the benefits of AI without sacrificing the determinism required for engineering applications~\cite{safety_net_ai_2023}.

\section{Conclusion}

\textbf{COOL} (Chain-Oriented Objective Logic) provides a neuro-symbolic framework for dynamic orchestration in rule-driven industrial systems. Integrating \textbf{Chain-of-Logic (CoL)} for divide-and-conquer and \textbf{Neural Network Feedback Control (NNFC)} with \textbf{cascade filtering}, the framework bridges neural adaptability and symbolic rigor. Formal analysis establishes complexity reduction and Lyapunov stability, while empirical evaluations on rule-intensive tasks demonstrate deterministic reliability (100\% accuracy) and a 90\% reduction in computational overhead. 

The "small but many" modular architecture ensures resilience against environmental drift and catastrophic forgetting in high-stakes engineering. \textbf{COOL} offers a scalable, interpretable, and efficient foundation for deterministic strategy synthesis in industrial control and autonomous reasoning environments.

\section*{Declaration of Generative AI and AI-assisted technologies in the writing process}
During the preparation of this work the author(s) used \textbf{Gemini} in order to \textbf{improve language clarity, polish academic expression, and optimize the manuscript's LaTeX formatting for the journal requirements}. After using this tool/service, the author(s) reviewed and edited the content as needed and take(s) full responsibility for the content of the publication.

\clearpage
\bibliography{references}

@article{gulwani2017program,
  title={Program synthesis},
  author={Gulwani, Sumit and Polozov, Oleksandr and Singh, Rishabh and others},
  journal={Foundations and Trends{\textregistered} in Programming Languages},
  volume={4},
  number={1-2},
  pages={1--119},
  year={2017},
  publisher={Now Publishers, Inc.}
}

@article{achiam2023gpt,
  title={Gpt-4 technical report},
  author={Achiam, Josh and Adler, Steven and Agarwal, Sandhini and Ahmad, Lama and Akkaya, Ilge and Aleman, Florencia Leoni and Almeida, Diogo and Altenschmidt, Janko and Altman, Sam and Anadkat, Shyamal and others},
  journal={arXiv preprint arXiv:2303.08774},
  year={2023}
}

@book{lesk1975lex,
  title={Lex: A lexical analyzer generator},
  author={Lesk, Michael E and Schmidt, Eric},
  volume={39},
  year={1975},
  publisher={Bell Laboratories Murray Hill, NJ}
}

@book{johnson1975yacc,
  title={Yacc: Yet another compiler-compiler},
  author={Johnson, Stephen C and others},
  volume={32},
  year={1975},
  publisher={Bell Laboratories Murray Hill, NJ}
}

@article{sujeeth2014delite,
  title={Delite: A compiler architecture for performance-oriented embedded domain-specific languages},
  author={Sujeeth, Arvind K and Brown, Kevin J and Lee, Hyoukjoong and Rompf, Tiark and Chafi, Hassan and Odersky, Martin and Olukotun, Kunle},
  journal={ACM Transactions on Embedded Computing Systems (TECS)},
  volume={13},
  number={4s},
  pages={1--25},
  year={2014},
  publisher={ACM New York, NY, USA}
}

@article{velickovic2017graph,
  title={Graph attention networks},
  author={Velickovic, Petar and Cucurull, Guillem and Casanova, Arantxa and Romero, Adriana and Lio, Pietro and Bengio, Yoshua and others},
  journal={stat},
  volume={1050},
  number={20},
  pages={10--48550},
  year={2017}
}

@article{hrinchuk2019tensorized,
  title={Tensorized embedding layers for efficient model compression},
  author={Hrinchuk, Oleksii and Khrulkov, Valentin and Mirvakhabova, Leyla and Orlova, Elena and Oseledets, Ivan},
  journal={arXiv preprint arXiv:1901.10787},
  year={2019}
}

@article{huang2015bidirectional,
  title={Bidirectional LSTM-CRF models for sequence tagging},
  author={Huang, Zhiheng and Xu, Wei and Yu, Kai},
  journal={arXiv preprint arXiv:1508.01991},
  year={2015}
}

@article{hart1968formal,
  title={A formal basis for the heuristic determination of minimum cost paths},
  author={Hart, Peter E and Nilsson, Nils J and Raphael, Bertram},
  journal={IEEE transactions on Systems Science and Cybernetics},
  volume={4},
  number={2},
  pages={100--107},
  year={1968},
  publisher={IEEE}
}

@inproceedings{wu2022graph,
  title={Graph neural networks: foundation, frontiers and applications},
  author={Wu, Lingfei and Cui, Peng and Pei, Jian and Zhao, Liang and Guo, Xiaojie},
  booktitle={Proceedings of the 28th ACM SIGKDD Conference on Knowledge Discovery and Data Mining},
  pages={4840--4841},
  year={2022}
}

@article{nye2020learning,
  title={Learning compositional rules via neural program synthesis},
  author={Nye, Maxwell and Solar-Lezama, Armando and Tenenbaum, Josh and Lake, Brenden M},
  journal={Advances in Neural Information Processing Systems},
  volume={33},
  pages={10832--10842},
  year={2020}
}

@article{sinha2019clutrr,
  title={CLUTRR: A diagnostic benchmark for inductive reasoning from text},
  author={Sinha, Koustuv and Sodhani, Shagun and Dong, Jin and Pineau, Joelle and Hamilton, William L},
  journal={arXiv preprint arXiv:1908.06177},
  year={2019}
}

@inproceedings{dsl_robotic_survey_2014,
  title={A survey on domain-specific languages in robotics},
  author={Nordmann, Arne and Hochgeschwender, Nico and Wrede, Sebastian},
  booktitle={International conference on simulation, modeling, and programming for autonomous robots},
  pages={195--206},
  year={2014},
  organization={Springer}
}

@article{dsl_robotic_design_2020,
  title={Software product line engineering for robotics},
  author={Brugali, Davide},
  journal={Software Engineering for Robotics},
  pages={1--28},
  year={2020},
  publisher={Springer}
}

@inproceedings{dsl_finance_2023,
  title={When Blockchain Meets Domain Specific Language: A Review},
  author={Liu, Chuan and Li, Jun and Lei, Hong and Xu, Xiang and Liu, Chao},
  booktitle={International Conference on Information Science, Communication and Computing},
  pages={110--125},
  year={2023},
  organization={Springer}
}

@article{dsl_challengeing_modularization,
  title={Challenges and Opportunities of Modularizing Textual Domain-Specific},
  author={Rieger, Christoph and Westerkamp, Martin and Kuchen, Herbert},
  journal={Model-Driven Software Development},
  pages={395},
  year={2018}
}

@article{dsl_modularization_summary_2005,
  title={When and how to develop domain-specific languages},
  author={Mernik, Marjan and Heering, Jan and Sloane, Anthony M},
  journal={ACM computing surveys (CSUR)},
  volume={37},
  number={4},
  pages={316--344},
  year={2005},
  publisher={ACM New York, NY, USA}
}

@inproceedings{dsl_modularization_challenge_2013,
  title={The state of the art in language workbenches: Conclusions from the language workbench challenge},
  author={Erdweg, Sebastian and Van Der Storm, Tijs and V{\"o}lter, Markus and Boersma, Meinte and Bosman, Remi and Cook, William R and Gerritsen, Albert and Hulshout, Angelo and Kelly, Steven and Loh, Alex and others},
  booktitle={International Conference on Software Language Engineering},
  pages={197--217},
  year={2013},
  organization={Springer}
}

@article{dsl_modularization_example_2024,
  title={Exploring the Effectiveness and Trends of Domain-Specific Model Driven Engineering: A Systematic Literature Review (SLR)},
  author={Zafar, Amina and Azam, Farooque and Latif, Afshan and Anwar, Muhammad Waseem and Safdar, Aon},
  journal={IEEE Access},
  volume={12},
  pages={86809--86830},
  year={2024},
  publisher={IEEE}
}

@inproceedings{dsl_modularization_exammple_2014,
  title={Domain specific languages for managing feature models: Advances and challenges},
  author={Collet, Philippe},
  booktitle={International Symposium On Leveraging Applications of Formal Methods, Verification and Validation},
  pages={273--288},
  year={2014},
  organization={Springer}
}

@inproceedings{neural_adaptive_control_robot_2025,
  title={Hybrid Fault-Tolerant Control in Cooperative Robotics: Advances in Resilience and Scalability},
  author={Urrea, Claudio},
  booktitle={Actuators},
  volume={14},
  number={4},
  pages={177},
  year={2025},
  organization={MDPI}
}

@article{neural_adaptive_control_reinforcement_2012,
  title={Reinforcement learning and optimal adaptive control: An overview and implementation examples},
  author={Khan, Said G and Herrmann, Guido and Lewis, Frank L and Pipe, Tony and Melhuish, Chris},
  journal={Annual reviews in control},
  volume={36},
  number={1},
  pages={42--59},
  year={2012},
  publisher={Elsevier}
}

@article{neurall_adaptive_feedback_control_2025,
  title={Control structures and algorithms for force feedback bilateral teleoperation systems: a comprehensive review},
  author={Tian, Jiawei and Zhou, Yu and Yin, Lirong and Alqahtani, Salman and Tang, Minyi and Lu, Siyu and Wang, Ruiyang and Zheng, Wenfeng},
  journal={Computer Modeling in Engineering \& Sciences},
  volume={142},
  number={2},
  pages={973},
  year={2025},
  publisher={Tech Science Press}
}

@article{heuristic_search_overview_2004,
  title={An overview of heuristic solution methods},
  author={Silver, Edward Allen},
  journal={Journal of the operational research society},
  volume={55},
  number={9},
  pages={936--956},
  year={2004},
  publisher={Taylor \& Francis}
}

@article{heuristic_search_cost_1989,
  title={New approaches for heuristic search: A bilateral linkage with artificial intelligence},
  author={Glover, Fred and Greenberg, Harvey J},
  journal={European Journal of Operational Research},
  volume={39},
  number={2},
  pages={119--130},
  year={1989},
  publisher={Elsevier}
}

@article{heuristic_search_weight_2017,
  title={Heuristic search: The emerging science of problem solving},
  author={Salhi, Sa{\"\i}d},
  year={2017},
  publisher={Springer}
}

@inproceedings{neural_control_system_stability_analysis_method,
  title={Neural networks for control: a tutorial and survey of stability-analysis methods, properties, and discussions},
  author={Norris, Griffin and Ducard, Guillaume JJ and Onder, Christopher},
  booktitle={2021 International Conference on Electrical, Computer, Communications and Mechatronics Engineering (ICECCME)},
  pages={1--6},
  year={2021},
  organization={IEEE}
}

@article{industrial_dsl_2022,
  title={Modeling languages in Industry 4.0: an extended systematic mapping study},
  author={Wortmann, Andreas and Barais, Olivier and Combemale, Benoit and Wimmer, Manuel},
  journal={Software and Systems Modeling},
  volume={19},
  number={1},
  pages={67--94},
  year={2020},
  publisher={Springer}
}

@article{deterministic_ai_2023,
  title={Surveying neuro-symbolic approaches for reliable artificial intelligence of things},
  author={Lu, Zhen and Afridi, Imran and Kang, Hong Jin and Ruchkin, Ivan and Zheng, Xi},
  journal={Journal of Reliable Intelligent Environments},
  volume={10},
  number={3},
  pages={257--279},
  year={2024},
  publisher={Springer}
}

@article{modular_edge_ai_2024,
  title={IoT and edge computing for architects},
  author={Lea, Perry},
  year={2020},
  publisher={Packt Publishing}
}

@article{dynamic_neural_search_2021,
  title={Enhancing dynamic symbolic execution by automatically learning search heuristics},
  author={Cha, Sooyoung and Hong, Seongjoon and Bak, Jiseong and Kim, Jingyoung and Lee, Junhee and Oh, Hakjoo},
  journal={IEEE Transactions on Software Engineering},
  volume={48},
  number={9},
  pages={3640--3663},
  year={2021},
  publisher={IEEE}
}

@article{modular_knowledge_2021,
  title={Application of knowledge based engineering principles to intelligent automation systems},
  author={Van der Velden, Christian},
  year={2024},
  school={RMIT University}
}

@article{modular_industrial_control_2023,
  title={Collective intelligence in self-organized industrial cyber-physical systems},
  author={Leit{\~a}o, Paulo and Queiroz, Jonas and Sakurada, Lucas},
  journal={Electronics},
  volume={11},
  number={19},
  pages={3213},
  year={2022},
  publisher={MDPI}
}

@article{reliable_industrial_reasoning_2023,
  title={Soft error tolerant convolutional neural networks on FPGAs with ensemble learning},
  author={Gao, Zhen and Zhang, Han and Yao, Yi and Xiao, Jiajun and Zeng, Shulin and Ge, Guangjun and Wang, Yu and Ullah, Anees and Reviriego, Pedro},
  journal={IEEE Transactions on Very Large Scale Integration (VLSI) Systems},
  volume={30},
  number={3},
  pages={291--302},
  year={2022},
  publisher={IEEE}
}

@article{eaai_neurosymbolic_2022,
  title={Cascade convolutional neural network with progressive optimization for motor fault diagnosis under nonstationary conditions},
  author={Wang, Fei and Liu, Ruonan and Hu, Qinghua and Chen, Xuefeng},
  journal={IEEE Transactions on Industrial Informatics},
  volume={17},
  number={4},
  pages={2511--2521},
  year={2020},
  publisher={IEEE}
}

@article{conflict_detection_2022,
  title={Conflict detection of functional requirements based on clustering and rule-based system},
  author={Yuhana, Umi Laili and Rochimah, Siti and others},
  journal={IEEE Access},
  volume={12},
  pages={174330--174342},
  year={2024},
  publisher={IEEE}
}

@article{maintenance_debt_2022,
  title={Automated Detection of Self-Admitted Technical Debt: A Systematic Literature Review},
  author={Abdulaziz, Anas and Sonbol, Riad and Alnoukari, Mouhib},
  journal={IEEE Access},
  year={2026},
  publisher={IEEE}
}

@article{industrial_ai_transparency_2021,
  title={Challenges with developing and deploying AI models and applications in industrial systems},
  author={Sinha, Sudhi and Lee, Young M},
  journal={Discover Artificial Intelligence},
  volume={4},
  number={1},
  pages={55},
  year={2024},
  publisher={Springer}
}

@inproceedings{adaptive_neurosymbolic_2022,
  title={Neuro-Symbolic AI for Self-Learning Intelligent Systems in Real-Time Industrial Automation},
  author={Madhura, K and Al-Assal, Jawad Radhi Rustum and Hussein, Omar Mohsen and Ibragimova, Kamila and Sajiv, G and Bala, B Kiran},
  booktitle={2025 IEEE Pune Section International Conference (PuneCon)},
  pages={1--5},
  year={2025},
  organization={IEEE}
}

@article{dynamic_planning_2023,
  title={HEADER: Hierarchical Robot Exploration via Attention-Based Deep Reinforcement Learning with Expert-Guided Reward},
  author={Cao, Yuhong and Wang, Yizhuo and Liang, Jingsong and Liao, Shuhao and Zhang, Yifeng and Li, Peizhuo and Sartoretti, Guillaume},
  journal={arXiv preprint arXiv:2510.15679},
  year={2025}
}

@article{cascade_verification_2023,
  title={Certified neural network control architectures: Methodological advances in stability, robustness, and cross-domain applications},
  author={Liu, Rui and Huang, Jianhua and Lu, Biao and Ding, Weili},
  journal={Mathematics},
  volume={13},
  number={10},
  pages={1677},
  year={2025},
  publisher={MDPI}
}

@article{neurosymbolic_review_2022,
  title={Neuro-symbolic artificial intelligence: a survey},
  author={Bhuyan, Bikram Pratim and Ramdane-Cherif, Amar and Tomar, Ravi and Singh, TP},
  journal={Neural Computing and Applications},
  volume={36},
  number={21},
  pages={12809--12844},
  year={2024},
  publisher={Springer}
}

@article{state_space_reduction_2021,
  title={Rule-based reinforcement learning for efficient robot navigation with space reduction},
  author={Zhu, Yuanyang and Wang, Zhi and Chen, Chunlin and Dong, Daoyi},
  journal={IEEE/ASME Transactions on Mechatronics},
  volume={27},
  number={2},
  pages={846--857},
  year={2021},
  publisher={IEEE}
}

@inproceedings{lyapunov_neural_control_2021,
  title={Neural networks for control: A tutorial and survey of stability-analysis methods, properties, and discussions},
  author={Norris, Griffin and Ducard, Guillaume JJ and Onder, Christopher},
  booktitle={2021 International Conference on Electrical, Computer, Communications and Mechatronics Engineering (ICECCME)},
  pages={1--6},
  year={2021},
  organization={IEEE}
}

@article{eaai_stability_2022,
  title={Neuro-Symbolic Predictive Process Monitoring},
  author={Mezini, Axel and Umili, Elena and Donadello, Ivan and Maggi, Fabrizio Maria and Mancanelli, Matteo and Patrizi, Fabio},
  journal={arXiv preprint arXiv:2509.00834},
  year={2025}
}

@article{symbolic_validation_2023,
  title={Neuro-symbolic artificial intelligence: Towards improving the reasoning abilities of large language models},
  author={Yang, Xiao-Wen and Shao, Jie-Jing and Guo, Lan-Zhe and Zhang, Bo-Wen and Zhou, Zhi and Jia, Lin-Han and Dai, Wang-Zhou and Li, Yu-Feng},
  journal={arXiv preprint arXiv:2508.13678},
  year={2025}
}

@article{formal_logic_synthesis_2021,
  title={Neural-symbolic learning and reasoning: A survey and interpretation},
  author={Hitzler, P and Sarker, MK and Besold, TR and Garcez, AD and Bader, S and Bowman, H and Domingos, P and Hitzler, P and K{\"u}hnberger, KU and Lamb, LC and others},
  journal={Frontiers in artificial intelligence and applications},
  volume={342},
  pages={1--51},
  year={2022}
}

@article{modular_complexity_2020,
  title={A decomposition-based development method for industrial control systems},
  author={Xiong, Jiawen and Li, Ju and Shi, Jianqi and Huang, Yanhong},
  journal={IEEE Access},
  volume={7},
  pages={93161--93174},
  year={2019},
  publisher={IEEE}
}

@book{expert_heuristics_2017,
  title={Heuristic Search: The Emerging Science of Problem Solving},
  author={Salhi, Sa{\"\i}d},
  year={2017},
  publisher={Springer},
  note={Theoretical foundation for knowledge-guided optimization.}
}

@inproceedings{search_space_pruning_2021,
  title={Efficient state space pruning in symbolic backward traversal},
  author={Cabodi, Gianpiero and Camurati, Paolo and Quer, Stefano},
  booktitle={Proceedings 1994 IEEE International Conference on Computer Design: VLSI in Computers and Processors},
  pages={230--235},
  year={1994},
  organization={IEEE}
}

@book{discrete_event_systems_2021,
  title={Introduction to discrete event systems},
  author={Cassandras, Christos G and Lafortune, Stephane},
  year={2007},
  publisher={Springer}
}

@article{modular_neural_networks_2021,
  title={A review of modularization techniques in artificial neural networks},
  author={Amer, Mohammed and Maul, Tom{\'a}s},
  journal={Artificial Intelligence Review},
  volume={52},
  number={1},
  pages={527--561},
  year={2019},
  publisher={Springer}
}

@article{informed_machine_learning_2021,
  title={Informed machine learning--a taxonomy and survey of integrating prior knowledge into learning systems},
  author={Von Rueden, Laura and Mayer, Sebastian and Beckh, Katharina and Georgiev, Bogdan and Giesselbach, Sven and Heese, Raoul and Kirsch, Birgit and Pfrommer, Julius and Pick, Annika and Ramamurthy, Rajkumar and others},
  journal={IEEE Transactions on Knowledge and Data Engineering},
  volume={35},
  number={1},
  pages={614--633},
  year={2021},
  publisher={IEEE}
}

@article{neurosymbolic_determinism_2023,
  title={Surveying neuro-symbolic approaches for reliable artificial intelligence of things},
  author={Lu, Zhen and Afridi, Imran and Kang, Hong Jin and Ruchkin, Ivan and Zheng, Xi},
  journal={Journal of Reliable Intelligent Environments},
  volume={10},
  number={3},
  pages={257--279},
  year={2024},
  publisher={Springer}
}

@article{cascaded_reliability_2022,
  title={Mind the gap: A causal perspective on bias amplification in prediction \& decision-making},
  author={Plecko, Drago and Bareinboim, Elias},
  journal={Advances in Neural Information Processing Systems},
  volume={37},
  pages={84384--84408},
  year={2024}
}

@article{consensus_learning_2023,
  title={Exploring various neural network configurations for the NN-based MPC in Multiagent System},
  author={Chaubey, Piyush and Markana, Anilkumar and Vyas, Dhaval and Goyal, Deepak Kumar},
  journal={Future Technology},
  volume={5},
  number={1},
  pages={148--158},
  year={2026}
}

@article{multitask_reasoning_2022,
  title={Efficient multi-task reinforcement learning with cross-task policy guidance},
  author={He, Jinmin and Li, Kai and Zang, Yifan and Fu, Haobo and Fu, Qiang and Xing, Junliang and Cheng, Jian},
  journal={Advances in Neural Information Processing Systems},
  volume={37},
  pages={117997--118024},
  year={2024}
}

@article{modular_consistency_2022,
  title={Composition of modular models for verification of distributed automation systems},
  author={Zeller, Andreas and Weyrich, Michael},
  journal={Procedia Manufacturing},
  volume={17},
  pages={870--877},
  year={2018},
  publisher={Elsevier}
}

@article{compositional_expressiveness_2021,
  title={Compositional embeddings of domain-specific languages},
  author={Sun, Yaozhu and Dhandhania, Utkarsh and Oliveira, Bruno C d S},
  journal={Proceedings of the ACM on Programming Languages},
  volume={6},
  number={OOPSLA2},
  pages={175--203},
  year={2022},
  publisher={ACM New York, NY, USA}
}

@article{adaptive_formal_systems_2023,
  title={A survey on context-aware multi-agent systems: techniques, challenges and future directions},
  author={Du, Hung and Thudumu, Srikanth and Vasa, Rajesh and Mouzakis, Kon},
  journal={arXiv preprint arXiv:2402.01968},
  year={2024}
}

@article{synthesis_complexity_2021,
  title={On the hardness of approximate reasoning},
  author={Roth, Dan},
  journal={Artificial intelligence},
  volume={82},
  number={1-2},
  pages={273--302},
  year={1996},
  publisher={Elsevier}
}

@article{search_tractability_2023,
  title={An adaptive hyperbox algorithm for high-dimensional discrete optimization via simulation problems},
  author={Xu, Jie and Nelson, Barry L and Hong, L Jeff},
  journal={INFORMS Journal on Computing},
  volume={25},
  number={1},
  pages={133--146},
  year={2013},
  publisher={INFORMS}
}

@article{informed_ml_2021,
  title={Informed machine learning--a taxonomy and survey of integrating prior knowledge into learning systems},
  author={Von Rueden, Laura and Mayer, Sebastian and Beckh, Katharina and Georgiev, Bogdan and Giesselbach, Sven and Heese, Raoul and Kirsch, Birgit and Pfrommer, Julius and Pick, Annika and Ramamurthy, Rajkumar and others},
  journal={IEEE Transactions on Knowledge and Data Engineering},
  volume={35},
  number={1},
  pages={614--633},
  year={2021},
  publisher={IEEE}
}

@book{fault_detection_2006,
  title={A review on fault detection and diagnosis techniques: basics and beyond},
  author={Abid, Anam and Khan, Muhammad Tahir and Iqbal, Javaid},
  journal={Artificial Intelligence Review},
  volume={54},
  number={5},
  pages={3639--3664},
  year={2021},
  publisher={Springer}
}

@article{state_space_mitigation_2023,
  title={Handling state space explosion in component-based software verification: A review},
  author={Nejati, Faranak and Abd Ghani, Abdul Azim and Yap, Ng Keng and Jafaar, Azmi Bin},
  journal={IEEE Access},
  volume={9},
  pages={77526--77544},
  year={2021},
  publisher={IEEE}
}

@inproceedings{modular_decomposition_2022,
  title={Model-based DSL frameworks},
  author={Kurtev, Ivan and B{\'e}zivin, Jean and Jouault, Fr{\'e}d{\'e}ric and Valduriez, Patrick},
  booktitle={Companion to the 21st ACM SIGPLAN symposium on Object-oriented programming systems, languages, and applications},
  pages={602--616},
  year={2006}
}

@article{neurosymbolic_search_2021,
  title={Fast task planning with neuro-symbolic relaxation},
  author={Du, Qiwei and Li, Bowen and Du, Yi and Su, Shaoshu and Fu, Taimeng and Zhan, Zitong and Zhao, Zhipeng and Wang, Chen},
  journal={IEEE Robotics and Automation Letters},
  year={2026},
  publisher={IEEE}
}

@article{concept_drift_review_2020,
  title={Learning under concept drift: A review},
  author={Lu, Jie and Liu, Anjin and Dong, Fan and Gu, Feng and Gama, Joao and Zhang, Guangquan},
  journal={IEEE Transactions on Knowledge and Data Engineering},
  volume={31},
  number={12},
  pages={2346--2363},
  year={2020},
  publisher={IEEE},
  note={Supports the motivation of non-stationary environment and drift.}
}

@article{dynamic_stability_2022,
  title={Dynamic learning from adaptive neural network control of a class of nonaffine nonlinear systems},
  author={Dai, Shi-Lu and Wang, Cong and Wang, Min},
  journal={IEEE transactions on neural networks and learning systems},
  volume={25},
  number={1},
  pages={111--123},
  year={2013},
  publisher={IEEE}
}

@inproceedings{search_compression_2022,
  title={Symbolic time and space tradeoffs for probabilistic verification},
  author={Chatterjee, Krishnendu and Dvo{\v{r}}{\'a}k, Wolfgang and Henzinger, Monika and Svozil, Alexander},
  booktitle={2021 36th Annual ACM/IEEE Symposium on Logic in Computer Science (LICS)},
  pages={1--13},
  year={2021},
  organization={IEEE}
}

@article{sparse_learning_2021,
  title={Alleviating data sparsity to enhance AI models robustness in IoT network security context},
  author={Sood, Keshav and Liu, Shigang and Nguyen, Dinh Duc Nha and Kumar, Neeraj and Feng, Bohao and Yu, Shui},
  journal={IEEE Transactions on Mobile Computing},
  volume={24},
  number={5},
  pages={3764--3778},
  year={2025},
  publisher={IEEE}
}

@article{continual_learning_industrial_2023,
  title={A continual learning approach for failure prediction under non-stationary conditions: Application to condition monitoring data streams},
  author={Benatia, Mohamed Amin and Hafsi, Meriem and Ayed, S Ben},
  journal={Computers \& Industrial Engineering},
  volume={204},
  pages={111049},
  year={2025},
  publisher={Elsevier}
}

@article{safety_net_ai_2023,
  title={Semantically-Aligned Reasoning for Safe AGI via Knowledge-Based Deep Models},
  author={Fatima, Sundas and others},
  journal={Journal of Computational Science and Applications},
  volume={2},
  number={1},
  pages={22--26},
  year={2025}
}
\bibliographystyle{ieeetr}

\clearpage
\appendix

\section{Detailed Theoretical Analysis of Chain-of-Logic}

\subsection{Expressiveness Preservation Under Finite-State Orchestration}
\label{app:prf_expressiveness}

\begin{proof}
Consider a Chain-of-Logic (CoL) framework orchestrating a finite set of sub-DSLs $\mathcal{G} = \{G_1, G_2, \dots, G_k\}$. We prove that the introduction of CoL regulation does not alter the fundamental computational expressiveness of the underlying system.

\textbf{System Definition:} Let the CoL controller be defined by a finite set of control states $Q_c = \{A_1, A_2, \ldots, A_m\}$, representing the expert-guided activity workflow. The state of the composite system at any time $t$ can be uniquely represented by the tuple $(q_c, q_d, s)$, where:
\begin{itemize}
    \item $q_c \in Q_c$ is the active activity (control state) within the CoL workflow.
    \item $q_d$ is the internal state of the currently active sub-DSL $G_{\text{active}} \in \mathcal{G}$.
    \item $s$ denotes the auxiliary storage (e.g., stack or tape) utilized by $G_{\text{active}}$.
\end{itemize}

\textbf{Upper Bound Analysis:} By definition, the CoL controller $Q_c$ is a finite-state transition system. The transition function $\delta_c$ of CoL only influences the selection of the active sub-DSL and the application of its rules based on the finite heuristic vectors and keywords. Crucially, CoL does not possess its own auxiliary storage (e.g., it does not add an additional stack to a pushdown automaton). Consequently, the state space of the composite system is a Cartesian product of a finite set $Q_c$ and the state space of $\mathcal{G}$. In the hierarchy of formal languages, the product of a finite-state automaton and a grammar $G$ remains within the same class of expressiveness as $G$. Thus:
\begin{equation}
\text{Expressiveness}(\text{CoL} + \mathcal{G}) \leq \max_{G_i \in \mathcal{G}} (\text{Expressiveness}(G_i))
\end{equation}

\textbf{Lower Bound Analysis:} Conversely, a CoL framework can be trivially configured to behave as a single DSL $G \in \mathcal{G}$ by defining a single control state $q_c$ that unconditionally invokes $G$ and remains in that state. This configuration simulates $G$ with zero regulatory overhead, implying:
\begin{equation}
\text{Expressiveness}(G) \leq \text{Expressiveness}(\text{CoL} + \mathcal{G})
\end{equation}

\textbf{Conclusion:} Combining the upper and lower bounds, we conclude that:
\begin{equation}
\text{Expressiveness}(\text{CoL} + \mathcal{G}) = \max_{G_i \in \mathcal{G}} (\text{Expressiveness}(G_i))
\end{equation}
This equality confirms that the finite-state orchestration of CoL acts as a transparent orchestrator. It ensures that COOL can be applied to any domain logic without restricting its inherent capability to solve complex engineering problems, while providing the necessary structure to suppress unregulated state-space explosion.
\end{proof}

\subsection{Formal Proof of CoL Regulatory Expressiveness}
\label{app:prf_regulatory_scaling}

To analyze the limits of CoL's coordination, we define its transition mechanism as a state-aware scheduling function.

\begin{definition}[CoL Control Function]
The CoL control function $\delta_c$ maps the current regulatory state to the next activity based on keyword directives and the operational state output from the active DSL:
\begin{equation}
    \delta_c(c_i, K, \sigma) = 
\begin{cases} 
c_{i + l(\sigma)} & \text{if } K = \mathtt{return} \\
c_{n(\sigma)} & \text{if } K = \mathtt{logicjump}(n) \\
\emptyset & \text{if } K = \mathtt{abort}
\end{cases}
\end{equation}
where $c_i \in Q_c$ is the current CoL activity, $K$ is the control primitive, and $\sigma$ denotes the operational state feedback from the governed DSL. The functions $l(\sigma)$ and $n(\sigma)$ represent the dynamic parameterization of the transition logic (where $n, i, l \in \mathbb{N}$).
\end{definition}

The computational expressiveness of the regulatory strategy $\delta_c$ is strictly bounded by the maximum expressiveness of the parameterization functions $n(\sigma)$ and $l(\sigma)$, which in turn depend on the feedback density of $\sigma$.

\begin{proof}
We establish the scaling properties of $\delta_c$ through a case-wise analysis of the computational hierarchy:

\textbf{Case 1: Finite-State Regulation.} 
If the governed DSLs generate finite-state outputs, then the state feedback is restricted to a finite alphabet $\sigma \in \Sigma_F$. Consequently, the mapping functions $n(\sigma)$ and $l(\sigma)$ possess finite ranges. In this scenario, $\delta_c$ functions as a \textit{finite-state transducer}, providing lightweight, deterministic scheduling for linear engineering workflows.

\textbf{Case 2: Context-Free Regulation.} 
If the underlying DSL provides stack-aware feedback (e.g., in nested recursive synthesis), $\sigma$ encodes the current stack configuration. The parameterization $n(\sigma)$ can thus implement stack-dependent transitions, enabling $\delta_c$ to orchestrate complex, hierarchical activities equivalent to a \textit{pushdown automaton's} control unit.

\textbf{Case 3: Turing-Complete Regulation.} 
If the DSL produces Turing-computable outputs (e.g., in open-ended program synthesis), $\sigma$ represents arbitrary tape states. Here, $n(\sigma)$ becomes a Turing-computable function, allowing $\delta_c$ to execute any logically computable regulatory strategy.

In all cases, the regulatory power of $\delta_c$ is not an intrinsic limitation of the COOL framework but is determined by the "sensor density" (operational feedback) of the governed DSLs. This proof confirms that CoL provides a flexible, adaptive control layer capable of managing engineering tasks of arbitrary complexity.
\end{proof}

\subsection{Formal Complexity Analysis of Chain-of-Logic}
\label{app:prf_complexity}

\subsubsection{Foundational Metrics and Baseline Complexity}

\begin{definition}[Search Complexity Metric]
The computational overhead of a symbolic synthesis process is characterized by the \textbf{branching factor} $b$ (the cardinality of the candidate rule set evaluated at each step) and the \textbf{derivation length} $L$ (the total number of rule applications). The cumulative search space is bounded by the exponential term $O(b^L)$.
\end{definition}

\begin{definition}[Monolithic Baseline Complexity]
In the absence of CoL regulation, the synthesizer must evaluate the entire global rule set $R$ at every step. The baseline complexity is defined as:
\begin{equation}
    \text{Complexity}_{\text{baseline}} = O\left( |R|^{L_{\text{global}}} \right)
\end{equation}
where $|R|$ denotes the size of the global rule set and $L_{\text{global}}$ is the path length under unguided, monolithic search conditions.
\end{definition}

\subsubsection{CoL State-Space Compression Operators}

The CoL framework introduces three orthogonal mechanisms that compress the search space independently of the underlying search algorithm:

\begin{lemma}[Sequential Activity Transition via \texttt{return}]
The \texttt{return} keyword triggers a deterministic increment of the activity index. This strictly reduces the active regulatory set from $\{R_i, \dots, R_N\}$ to $\{R_{i+1}, \dots, R_N\}$, inducing a monotonically non-increasing sequence of branching factors $b_t$.
\end{lemma}

\begin{lemma}[Non-sequential Routing via \texttt{logicjump(n)}]
The \texttt{logicjump(n)} keyword redefines the active search envelope by resetting the activity index to $n$:
\begin{itemize}
    \item \textbf{Exploration Concentration ($n > i$):} The active rule set $|R_{\text{active}}|$ is restricted to downstream activities, significantly reducing $b_t$.
    \item \textbf{Completeness Recovery ($n < i$):} The search envelope expands to include upstream rules only when necessary to preserve global reachability.
\end{itemize}
\end{lemma}

\begin{lemma}[Deterministic Branch Elimination via \texttt{abort}]
The \texttt{abort} keyword functions as an early failure detection operator. It introduces a pruning ratio $p_t$ ($0 \leq p_t < 1$) at step $t$, effectively reducing the local search volume from $b_t$ to $(1-p_t) \cdot b_t$.
\end{lemma}

\subsubsection{Composite Complexity and Asymptotic Bounds}

The synergistic effect of these operators yields the total regulated complexity:
\begin{equation}
    \text{Complexity}_{\text{CoL}} = O\left( \prod_{t=1}^{L} \left[ (1 - p_t) \cdot b_t \right] \right)
\end{equation}
where $b_t$ is the dynamically restricted branching factor, $p_t$ is the pruning effectiveness, and $L$ is the path length in the regulated space.

\textbf{Worst-Case Bound (Disabled Regulation):} If CoL transitions are inactive ($i=1$) and pruning fails ($p_t=0$), $b_t$ remains at $|R|$, and the complexity reverts to the monolithic baseline $O(|R|^L)$.

\textbf{Best-Case Bound (Perfect Regulation):} Under ideal conditions where expert heuristics perfectly restrict $b_t$ and \texttt{abort} prunes all irrelevant branches such that $(1-p_t)b_t = 1$, the product reduces to $1^L = 1$. Since $L$ steps must still be executed to complete the derivation, the lower bound is $\Omega(L)$.

\textbf{Conclusion:} These bounds prove that CoL offers a robust mechanism for suppressing exponential search growth. By compounding exponential branching restrictions with multiplicative pruning, CoL ensures that complex multi-DSL coordination remains computationally efficient and scalable for real-world engineering tasks.

\section{Theoretical Foundation of NNFC Integration}

\subsection{Proof of the Composability of Heuristics and Neural Oracles}
\label{app:prf_composability}

In complex engineering search tasks, we define the search process using heuristic values (e.g., cost, reward, or probability) and treat the neural agent as a \textit{Neural Oracle} providing real-time guidance. We prove that this integration maintains fundamental algorithmic properties while enhancing industrial performance.

\subsubsection{Composability under Completeness and Optimality}

\begin{proof}
We analyze the impact of neural guidance across four standard heuristic classes:

\begin{enumerate}
    \item \textbf{Cost Estimation (e.g., A*):} The neural oracle provides estimates of the remaining cost to the target. By ensuring the neural signal $\mathbf{\pi}_{\text{neural}}$ aligns with or underestimates the true cost (admissibility), the modified algorithm maintains optimality. Since the underlying symbolic transition rules are never excluded, search completeness is inherently preserved~\cite{heuristic_search_overview_2004,heuristic_search_cost_1989}.
    
    \item \textbf{Priority Weighting:} Neural signals adjust the prioritization of rule applications through multiplicative scaling. Because this only reorders the queue without pruning nodes, all valid derivation paths remain accessible, thus preserving completeness. Guidance toward the optimal path preserves system-wide optimality~\cite{heuristic_search_weight_2017}.
    
    \item \textbf{Probabilistic Selection:} The neural oracle increases the selection probability of promising rules. By enforcing non-zero probabilities for all authorized rules through renormalization, the system ensures that every part of the search space remains selectable, satisfying completeness requirements.
    
    \item \textbf{Reward Modulation:} Neural feedback adaptively tunes the rewards for specific rule-application actions. Completeness is maintained as long as the underlying action space remains unchanged, while optimality is preserved if reward updates favor globally optimal trajectories.
\end{enumerate}

\textbf{Failure Mode Analysis (Robustness):} If the neural oracle provides erroneous guidance (misprediction), optimality may be temporarily degraded (i.e., the system explores sub-optimal branches). However, since the CoL backbone maintains the full set of permissible rules, completeness is guaranteed unless the AI logic is allowed to explicitly delete valid solutions—a condition forbidden by the COOL architecture.
\end{proof}

\subsubsection{Composability under Engineering Resource Constraints}

In real-world engineering, where time and memory are finite, we evaluate the benefit of NNFC through the probability of success $P_{\text{found}}$ and expected time $E[T]$.

\begin{proof}
Let $\gamma$ be the probability of the oracle being active and $(1-\epsilon)$ be its accuracy. 
When the neural guidance is correct, it reduces the effective search volume by a factor $k \in (0, 1)$, increasing the success probability: $P'_{\text{found, correct}} = \frac{T}{k|S|}$. 
Even if the oracle is erroneous with rate $\epsilon$, the overall success probability is:
\begin{equation}
    P'_{\text{found}} = (1 - \epsilon) P'_{\text{found, correct}} + \epsilon P'_{\text{found, error}}
\end{equation}
For any reasonably trained neural agent where $\epsilon$ is low, $P'_{\text{found}} > P_{\text{found}}$ holds. 

Similarly, the expected execution time $E'[T]$ under hybrid guidance is:
\begin{equation}
    E'[T] = E[T] \left[ (1 - \gamma) + \gamma (1 - \epsilon) \beta + \gamma \epsilon \beta_{\text{error}} \right]
\end{equation}
where $\beta < 1$ represents the acceleration from correct guidance. This proves that the hybrid AI-expert approach significantly reduces the average computational cost of complex synthesis tasks compared to purely symbolic baselines.
\end{proof}

\textbf{Conclusion on Interdependency:} These proofs demonstrate that the reliability of the COOL framework is not solely dependent on AI accuracy. Instead, the final engineering outcome is protected by a \textbf{Harmonious Composability}: the symbolic CoL provides the safety "floor" (completeness), while the neural NNFC provides the performance "ceiling" (efficiency). This synergy is essential for deploying AI in critical engineering applications.

\subsection{Formal Analysis of Cascade Filtering}
\label{app:prf_cascade_filtering}

This section provides a rigorous mathematical analysis of the \textbf{cascade filtering mechanism} within the NNFC framework. We demonstrate how the series architecture enhances the system's ability to self-correct by transforming latent neural errors into detectable signal discrepancies.

\subsubsection{Error Transfer and Cumulative Amplification}

\begin{proof}
Consider a series of $K$ neural networks where each network $i$ aims to approximate a target function $f_i(\theta_i)$. The actual output $\pi_i$ is influenced by the input error from the previous stage $e_{i-1}$ and its own stochastic error term $\epsilon_i$ (with expectation $\epsilon$). Using a first-order Taylor approximation, the output of the $i$-th stage is:
\begin{equation}
    \pi_i = f_i(\theta_i + k_i \cdot e_{i-1}) + \epsilon_i \approx f_i(\theta_i) + \frac{\partial f_i}{\partial \theta_i} \cdot k_i \cdot e_{i-1} + \epsilon_i
\end{equation}
where $k_i \in [0,1]$ is a coupling coefficient. Defining the error transfer coefficient as $\gamma_i = \frac{\partial f_i}{\partial \theta_i} \cdot k_i$, the cumulative error $e_i$ at stage $i$ evolves as:
\begin{equation}
    e_i = \pi_i - f_i(\theta_i) \approx \gamma_i \cdot e_{i-1} + \epsilon_i
\end{equation}
Assuming identical error characteristics ($\epsilon_i = \epsilon, \gamma_i = \gamma$) across the cascade and an initial error $e_0 = 0$, the cumulative error at level $K$ follows a geometric progression:
\begin{equation}
    e_K \approx \epsilon \sum_{j=0}^{K-1} \gamma^j = \epsilon \frac{1 - \gamma^K}{1 - \gamma} \quad (\text{for } \gamma \neq 1)
\end{equation}
This derivation proves that for $\gamma \approx 1$, the cumulative error $e_K$ grows linearly with the depth $K$, effectively amplifying the signal discrepancy induced by neural mispredictions.
\end{proof}

\subsubsection{Sensitivity Comparison: Series vs. Parallel Structures}

To quantify filtering effectiveness, we define the \textbf{Inconsistency Metric ($D$)} as the normalized maximum discrepancy between network outputs: $D = \frac{\max_{i,j} \| \pi_i - \pi_j \|}{\| \pi_{\text{baseline}} \|}$.

\begin{proof}
We compare the cascade metric $D$ against a parallel ensemble baseline $D_{\text{avg}}$ (where outputs are averaged).
\begin{enumerate}
    \item \textbf{High-Sensitivity Regime (Low $\epsilon$):} When the neural agent is relatively accurate, the parallel ensemble tends to smooth out discrepancies, leading to $D_{\text{avg}} \propto \sqrt{\epsilon}$. In contrast, the series structure accumulates errors, ensuring $E[D] \propto e_K / \| \pi^* \|$. Since $e_K > \sqrt{\epsilon}$ for $K \geq 2$, the cascade structure exhibits higher sensitivity, making it a superior "tripwire" for detecting subtle logic deviations.
    \item \textbf{Robustness Regime (High $\epsilon$):} In scenarios with high neural uncertainty, the series structure may suffer from excessive variance ($D > D_{\text{avg}}$). However, by optimizing the cascade depth $K$ and the filtering threshold $T$, the system can maintain a superior signal-to-noise ratio, ensuring that high-error predictions are blocked more aggressively than in parallel setups.
\end{enumerate}
This analysis confirms that cascade provides an enhanced "filtering sensitivity" that is essential for maintaining the integrity of symbolic reasoning.
\end{proof}

\subsubsection{Adaptive Filtering Probability Law}

The final filtering probability $p_{\text{filter}}$ must adapt to environmental volatility.

\begin{proof}
Based on empirical observations, the network uncertainty $\gamma$ relates inversely to the error rate ($\gamma \propto 1 - \epsilon$). Substituting this into the cumulative error model, we obtain the adaptive filtering law:
\begin{equation}
    p_{\text{filter}} = P(D > T) \approx \beta \cdot D \propto \beta \cdot \epsilon \cdot \frac{1 - \gamma^K}{1 - \gamma}
\end{equation}
where $\beta$ is a proportionality constant. This relationship proves that the NNFC framework is \textbf{self-adaptive}: as the underlying neural error rate $\epsilon$ increases due to task complexity or adversarial conditions, the system's filtering probability $p_{\text{filter}}$ increases non-linearly, effectively shielding the Chain-of-Logic from erroneous AI guidance.
\end{proof}

\subsection{Formal Stability Analysis of the NNFC Framework}
\label{app:prf_lyapunov}

This section provides the mathematical modeling of neural error components and the formal Lyapunov-based proof of the system's adaptive stability under dynamic conditions.

\subsubsection{Mathematical Modeling of Neural Error Components}

In the COOL framework, each modular neural agent is modeled as a parametric function $\mathbf{\pi}(a | s; \theta)$ that generates rule-selection strategies. To analyze its reliability, we decompose the total operational error $e_{\text{total}}$ into static and dynamic components:
\begin{equation}
e_{\text{total}} = e_{\text{static}} + e_{\text{dynamic}} = e_{\text{static}} + (e_{\text{drift}} + e_{\text{forget}})
\end{equation}

\begin{definition}[Static Error $e_{\text{static}}$]
The static error represents the residual error after convergence, stemming from limited model capacity or data sparsity: $e_{\text{static}} = L(\mathbf{\pi}_{n-1}, \mathbf{\pi}^*_{n-1})$. In ideal convex optimizations, $e_{\text{static}} \to 0$, but in complex engineering tasks, it typically converges to a small non-zero constant.
\end{definition}

\begin{definition}[Drift Error $e_{\text{drift}}$]
Drift error arises from non-stationary environments (e.g., changes in DSL rule-sets): $e_{\text{drift}} = \alpha \| \mathbf{\pi}^*_n - \mathbf{\pi}^*_{n-1} \|$, where $\alpha$ is the sensitivity to environmental shifts.
\end{definition}

\begin{definition}[Forget Error $e_{\text{forget}}$]
Forget error occurs during continual learning as new synthesis trajectories overwrite historical knowledge: $e_{\text{forget}} = \gamma \sum_{i=1}^{n-1} \| \mathbf{\pi}_i - \mathbf{\pi}_{i-1} \|$, where $\gamma \propto \eta/C$ depends on the learning rate $\eta$ and network capacity $C$.
\end{definition}

\subsection{Formal Lyapunov Stability Analysis for NNFC}
\label{app:prf_lyapunov_full}

To prove the stability of the Neural Network Feedback Control (NNFC) under dynamic multi-DSL regulation, we model the system as a discrete-time adaptive controller and analyze its convergence properties.

\subsubsection{Energy Function Definition}
We define a positive-definite Lyapunov function $V$ as the sum of the neural approximation energy and the system output energy:
\begin{equation}
V = V_{\text{neural}} + V_{\text{output}} = \| \mathbf{\pi}^* - \mathbf{\pi}_{\text{neural}} \|^2 + e_{\text{output}}^2
\end{equation}
where $\mathbf{\pi}^*$ is the ideal regulatory strategy, $\mathbf{\pi}_{\text{neural}}$ is the neural agent's compensation, and $e_{\text{output}} = \| \mathbf{\pi}^* - \mathbf{\pi}_{\text{output}} \|$ is the final regulation error.

\subsubsection{Dynamics of Energy Dissipation and Knowledge Retention}

In rule-intensive engineering tasks, forgetting occurs as parameter updates for new derivation paths gradually overwrite previously optimized logical manifolds. We model the \textit{cumulative forgetting residual} at step $n$ as a path integral of the weight updates across the learning history. Furthermore, considering a continuous training cycle consisting of $l$ update steps, the energy change $\Delta V_{\text{neural}} = V_{\text{neural}, n+l} - V_{\text{neural}, n}$ must account for the net balance between gradient-based dissipation and the accumulation of dynamic disturbances. 

Combining the dissipative learning flow with the cumulative forgetting model, the energy evolution over a cycle of length $l$ is bounded by:
\begin{equation}
    \Delta V_{\text{neural}} \leq \sum_{i=n}^{n+l-1} \left( - \eta \| \nabla L(\theta_i) \|^2 \right) + \delta \cdot \underbrace{\left[ e_{\text{drift}} + \gamma \sum_{i=1}^{n+l} \eta \| \nabla L(\theta_{i-1}) \| \right]}_{e_{\text{dynamic}}} + O(\eta^2)
\end{equation}
where $\eta$ is the learning rate, and $e_{\text{dynamic}}$ represents the total disturbance composed of the environmental drift $e_{\text{drift}}$ and the integrated forgetting residual $e_{\text{forget}}$. The latter is explicitly parameterized by $\gamma$, which dictates the rate at which parameter-space movement erodes existing logical structures. Stability in the neural layer requires the gradient-based reduction to dominate the dynamic disturbance.

\textbf{2. Evolution of $V_{\text{output}}$:} 
The output error change $\Delta e_{\text{output}}$ is influenced by the adaptive learning efficiency $\alpha$ and the termination rate $\rho$, which is itself a function of the error $\rho = \kappa (1 - e_{\text{output}})$. The dynamics follow:
\begin{equation}
\Delta e_{\text{output}} \approx -\alpha \rho e_{\text{output}} + \beta e_{\text{dynamic}}
\end{equation}
The change in output energy is then derived via Taylor expansion:
\begin{equation}
\Delta V_{\text{output}} \approx 2 e_{\text{output}} \Delta e_{\text{output}} = 2 e_{\text{output}} \left( -\alpha \kappa (1 - e_{\text{output}}) e_{\text{output}} + \beta e_{\text{dynamic}} \right)
\end{equation}
Expanding the terms, we obtain:
\begin{equation}
\Delta V_{\text{output}} = -2\alpha \kappa (1 - e_{\text{output}}) e_{\text{output}}^2 + 2\beta e_{\text{dynamic}} e_{\text{output}}
\end{equation}

\subsubsection{Stability Conditions and Critical Threshold}

For global stability, we require the total energy to be non-increasing, $\Delta V \leq 0$. Concentrating on the output layer's stability:
\begin{equation}
-2\alpha \kappa (1 - e_{\text{output}}) e_{\text{output}}^2 + 2\beta e_{\text{dynamic}} e_{\text{output}} \leq 0
\end{equation}
Assuming $e_{\text{output}} > 0$, we divide by $2 e_{\text{output}}$ to obtain the stability constraint:
\begin{equation}
\beta e_{\text{dynamic}} \leq \alpha \kappa (1 - e_{\text{output}}) e_{\text{output}}
\end{equation}
Let $f(e_{\text{output}}) = \alpha \kappa (1 - e_{\text{output}}) e_{\text{output}}$. This quadratic function represents the system's "dissipative capacity" and achieves its maximum value at $e_{\text{output}} = 1/2$:
\begin{equation}
\max f(e_{\text{output}}) = \frac{\alpha \kappa}{4}
\end{equation}
Thus, the \textbf{Critical Stability Threshold} for the dynamic environment is:
\begin{equation}
e_{\text{dynamic}} \leq \frac{\alpha \kappa}{4\beta}
\end{equation}
If the dynamic error exceeds this threshold, the neural compensation cannot keep pace with environmental shifts, potentially leading to system divergence.

\subsubsection{Steady-State Error Derivation}

When the stability condition is satisfied, the system converges to a steady state where $\Delta e_{\text{output}} = 0$. Setting the dynamics equation to zero:
\begin{equation}
\alpha \kappa (1 - e_{\text{steady}}) e_{\text{steady}} = \beta e_{\text{dynamic}}
\end{equation}
Solving for $e_{\text{steady}}$ using the quadratic formula:
\begin{equation}
e_{\text{steady}} = \frac{1 - \sqrt{1 - \frac{4\beta e_{\text{dynamic}}}{\alpha \kappa}} }{2}
\end{equation}
For small $e_{\text{dynamic}}$ (typical in stable engineering processes), we apply a first-order Taylor approximation $\sqrt{1-x} \approx 1 - x/2$:
\begin{equation}
e_{\text{steady}} \approx \frac{1 - (1 - \frac{2\beta e_{\text{dynamic}}}{\alpha \kappa})}{2} = \frac{\beta e_{\text{dynamic}}}{\alpha \kappa}
\end{equation}
This derivation proves that the steady-state error of COOL is linearly proportional to the dynamic volatility and inversely proportional to the learning efficiency, providing a clear performance bound for industrial deployment.

\subsubsection{Avoidance of the Learning Singularity}
A critical failure mode in adaptive systems occurs when $e_{\text{output}} \to 1$, which causes the termination rate $\rho \to 0$, effectively freezing the learning process and allowing error to accumulate unchecked. The \textbf{cascade filtering structure} in NNFC ensures that erroneous neural signals are blocked before they can push $e_{\text{output}}$ into this singularity zone. By maintaining $\rho \geq \rho_{\text{min}} > 0$, the filter guarantees that the dissipative term in $\Delta V$ remains active, enabling the system to recover from high-error states.

\section{COOL Host Language: Design and Specification}
\label{app:cool_language}

The COOL framework is underpinned by a dedicated, high-performance execution engine and its corresponding host language, designed to meet the rigorous demands of neuro-symbolic regulation. To support complex engineering logic, the language employs a \textit{hybrid reasoning approach} featuring dynamic typing, full support for numerical operations, and random-access storage capabilities. This architecture allows the framework to maintain stateful read/write operations across heterogeneous DSL boundaries, providing the necessary expressiveness for large-scale coordination tasks. For detailed language specifications, syntax guides, and runtime documentation, please refer to the GitHub repository at \url{https://coolang.org}.

\begin{figure}[htbp]
    \centering
    \includegraphics[width=1\linewidth]{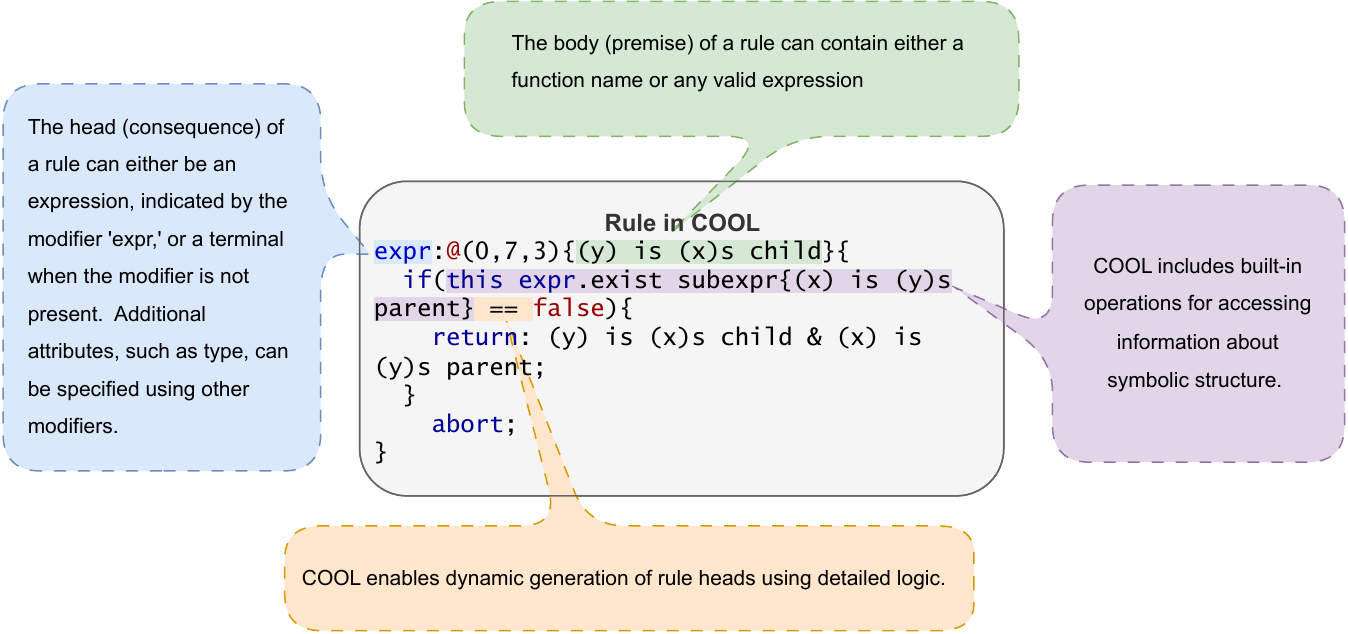}
    \caption{Architecture of DSL rules in the COOL environment. The engine supports defining rule heads via complex expressions or modifier-enhanced terminals. Rule bodies are designed for high extensibility, allowing for any valid logic expression or built-in function to access and manipulate the underlying symbolic structure dynamically.}
    \label{fig:cool_syntax}
\end{figure}

A representative COOL rule is illustrated in Figure~\ref{fig:cool_syntax}, showcasing the native integration of expert-defined heuristic vectors and runtime control keywords.

\subsection{Program Specification and Task Representation}

The following snippet demonstrates a typical \textit{Relational Reasoning} specification within the COOL environment:

\asumr{Relational Reasoning Task Input Program}{cool_code_family}{
\hspace{-5pt}\texttt{(Wesley) is (James)s son \textcolor{blue}{\&} (Martha) is (Wesley)s daughter \textcolor{blue}{\&} (Hugh) is (Martha)s uncle \textcolor{blue}{\&} (Hugh) is (James)s (\textcolor{olive}{\$relation});}
}

In this specification, the \verb|$| operator denotes a \textbf{nonterminal probe}, instructing the COOL engine to coordinate across multiple modular DSLs to resolve the correct value for the \verb|relation| variable. Rather than relying on a hard-coded solver, COOL facilitates a collaborative reasoning process where each sub-DSL identifies and solves localized constraints based on its internal domain logic. This process is orchestrated by CoL’s control primitives and adaptively optimized by NNFC’s feedback signals. All reasoning operations are executed via a unified compilation pipeline at the \textit{Intermediate Representation} level (detailed in Appendix~\ref{app:cool_ir}).

\subsection{Formal DSL Substrate and Terminology}
\label{app:dsl_term}

To provide a rigorous basis for the complexity and stability metrics used throughout this paper, we define the underlying DSL model as a standard context-free grammar (CFG):
\begin{equation}
    G = \{V, \Sigma, R, S\},
\end{equation}
where $V$ represents the set of non-terminal symbols, $\Sigma$ is the set of terminal symbols, $R$ is the set of production rules, and $S$ is the start symbol. The derivation process in COOL involves the iterative transformation of intermediate symbolic states through the application of rules within $R$ to generate complete, verifiable outputs.

During the derivation, each atomic step—consisting of an intermediate state ($s$) and its associated rule application ($r$)—forms a \textbf{transformation pair} $(s, r)$. A sequence of these pairs constitutes a \textbf{derivation trajectory}, which we categorize into three engineering states:
\begin{itemize}
    \item \textbf{Feasible Path}: A trajectory that successfully converges to a complete, valid solution.
    \item \textbf{Infeasible Path}: A trajectory that enters a terminal state without resolving the target nonterminals.
    \item \textbf{Unterminated Path}: A trajectory that remains in progress, typically constrained by current resource allocations.
\end{itemize} 

To quantify the computational cost of these processes, the following metrics are defined:
\begin{itemize}
    \item \textbf{Tree Operation (Ops)}: The physical modification of the abstract syntax tree during derivation, directly correlating to CPU utilization.
    \item \textbf{Transformation Pair $(s, r)$}: The fundamental unit of search space exploration, representing the memory-intensive state-rule mappings that the framework must track.
    \item \textbf{Derivation Path/Trajectory}: The complete execution trace from the initial problem specification to the final symbolic output, used to evaluate the end-to-end efficiency and stability of the regulation policy.
\end{itemize}

\subsection{A* Heuristic Search for Compiler-Level Program Synthesis}

To navigate the discrete state space of program synthesis effectively, COOL leverages the A* algorithm as its core search engine. A* is renowned for its efficacy in solving discrete optimization and path-finding tasks by utilizing heuristic guidance to steer the search toward high-probability regions~\cite{hart1968formal}. In the COOL framework, each rule-application decision is treated as an action with an associated cost. To align with the reward-based feedback from the neural agents, we interpret positive heuristic rewards as \textit{negative costs}. 

The search objective is to minimize the cumulative cost function $f(s) = g(s) + h(s)$, where $g(s)$ represents the cost incurred from the initial state and $h(s)$ is the estimated cost to reach a complete solution. Specifically, the dynamic compensation signal $u_2$ provided by the NNFC is integrated directly into the $g$-score and $f$-score calculations, allowing the neural oracle to prioritize promising derivation paths in real-time. Algorithm~\ref{alg:A_star} details the implementation of this neuro-symbolic search process.

\begin{algorithm}[h]
\caption{Search Algorithm in COOL}
\label{alg:A_star}
\begin{algorithmic}
\FUNCTION{ A* Search ($initialState, u_2$)}
    \STATE $openSet \gets \text{priority queue containing only the initial state}$
    \STATE $gScore[startState] \gets 0$ \COMMENT{cost from start}
    \STATE $fScore[startState] \gets 0$
    \WHILE{$openSet \neq \emptyset$}
        \STATE $currentState \gets openSet.\text{pop( ) }$ \COMMENT{State in openSet with lowest fScore value}
        \IF{$currentState \text{ is complete output}$}
            \STATE \textbf{return } \text{Success}
        \ENDIF
        \FORALL{$neighbor \in \text{neighbors of } currentState$}  \STATE \COMMENT{ Neighbor is obtained by applying a rule to the current state}
            \STATE $tentative\_gScore \gets gScore[current] - u_2[neighbor]$
            \IF{$tentative\_gScore < gScore[neighbor]$}
                \STATE $cameFrom[neighbor] \gets current$
                \STATE $gScore[neighbor] \gets tentative\_gScore$
                \STATE $fScore[neighbor] \gets gScore[neighbor] - u_2[neighbor]$
                \IF{$neighbor \not\in openSet$}
                    \STATE $openSet.\text{add}(neighbor)$
                \ENDIF
            \ENDIF
        \ENDFOR
    \ENDWHILE 
    \STATE \textbf{return } \text{Failure}
\ENDFUNCTION
\end{algorithmic}
\end{algorithm}

\subsection{Chain-of-Logic (CoL) Orchestration and State Refinement}
The coordination process within CoL is designed as a sequence of specialized logical activities, where each sub-DSL handles a specific functional aspect of the overall synthesis task. This modular approach allows for the progressive refinement of symbolic representations through structured, cross-DSL transformations. Crucially, intermediate sub-DSLs can generate \textit{partial symbolic states}, which serve as verified inputs for subsequent modules in the coordination chain, thereby preventing the logic of one domain from polluting the state space of another.

As illustrated in the relational reasoning example (Figure~\ref{fig:control_words}), the CoL pipeline partitions the global logic into four discrete activities:
\begin{itemize}
    \item \textbf{Entity Decomposition:} The first sub-DSL isolates basic relations and attributes (e.g., genders), simplifying the initial high-entropy representation into primitive components.
    \item \textbf{Inverse Logic Synthesis:} The second sub-DSL focuses exclusively on identifying and structuring inverse connections, narrowing the search to symmetry-based rules.
    \item \textbf{Transitive/Indirect Reasoning:} The third sub-DSL handles recursive and indirect relationships, utilizing contextual information extracted by previous stages to infer deep logical links.
    \item \textbf{Logic Recombination and Pruning:} The final sub-DSL reintegrates the refined relations while actively eliminating redundant or irrelevant data to produce a deterministic, complete solution.
\end{itemize}

Throughout this orchestrated execution, CoL utilizes expert-defined heuristic vectors to maintain high-level control, while runtime keywords enable flexible transitions between these specialized logic domains. This divide-and-conquer strategy is essential for maintaining computational tractability in large-scale, rule-intensive engineering regulation tasks.

\section{Neural Network Architecture and Training in COOL}
\label{app:neural}

The COOL framework incorporates an integrated automated learning system designed to extract regulatory knowledge from multi-DSL synthesis trajectories. This system orchestrates the data collection, feature embedding, and multi-objective training of the lightweight neural agents to achieve high-precision dynamic compensation.

\subsection{Data Acquisition and Multi-Head Dataset Construction}

To empower the neural agents with both task-awareness and search efficiency, COOL utilizes transformation pairs $(s, r)$ extracted from expert-guided derivation trajectories. For any regulation task of type $T$, the framework constructs a specialized dataset to train three distinct functional heads:

\textbf{Task Detection Head (TDH)}: This head functions as a contextual filter to determine if the input symbolic state resides within the operational manifold of task type $T$. In a binary classification setup, intermediate states from type $T$ trajectories serve as positive samples (67\%), while states from unrelated domains or built-in functions serve as negative samples (33\%) to suppress cross-task interference.

\textbf{Search Space Prune Head (SSPH)}: Conditioned on a positive TDH result, this head assesses the reachability of a complete solution from the current state. By training on feasible (67\%) and infeasible (33\%) derivation paths, the SSPH learns to identify "dead-end" branches, enabling aggressive pruning to preserve computational resources.

\textbf{Search Guidance Head (SGH)}: For feasible paths, the SGH generates high-level policy features. It employs a combination of classification and regression tasks to prioritize the most efficient production rules, effectively providing the dynamic priority compensation defined in the NNFC mechanism.

\begin{table}[!h]
\caption{Input feature specification for the cascaded neural agents in COOL. These features encode both the topological structure and the semantic context of the symbolic state, facilitating coordinated regulation across the CoL chain.}
\label{tab:input_features}

\centering
\renewcommand{\arraystretch}{1.2}
\begin{tabular}{p{0.13\textwidth}p{0.10\textwidth}p{0.10\textwidth}p{0.55\textwidth}}
\toprule
\centering \textbf{Feature} & \centering \textbf{Dim.} & \centering \textbf{Network} & \centering \textbf{Signification} \tabularnewline
\midrule
\centering grounded & \centering 2 & \centering A, B, C & Boolean flag indicating if the node belongs to a fully resolved symbolic expression. \tabularnewline

\centering domain & \centering 1 & \centering A, B, C & Encodes the task-specific activity context of the current subtree logic. \tabularnewline

\centering root & \centering 2 & \centering A, B, C & Identifies the entry node (root) of the subtask-specific derivation tree. \tabularnewline

\centering non-terminal & \centering 2 & \centering A, B, C & Indicates if the node is a symbolic non-terminal requiring further rule application. \tabularnewline

\centering type & \centering 1 & \centering A, B, C & Categorical encoding of the node's semantic and syntactic role. \tabularnewline

\centering identifier & \centering 1 & \centering A, B, C & Unique identifier mapping for the specific symbolic token. \tabularnewline

\centering string & \centering 1 & \centering A, B, C & Indicates the presence of a literal string value within the node. \tabularnewline

\centering number & \centering 1 & \centering A, B, C & Indicates the presence of a numerical constant within the node. \tabularnewline

\centering operator & \centering 1 & \centering A, B, C & Identifies the node as a functional operator in the symbolic expression. \tabularnewline

\centering current DSL & \centering 1 & \centering A, B, C & The active sub-DSL ID within the CoL chain (valid for grounded structures). \tabularnewline

\centering operand position & \centering 3  & \centering A, B, C & Structural encoding of node placement within binary syntax trees (Left, Right, or Parent). \tabularnewline

\centering applied (SGH) & \centering 1 & \centering B, C & Rule-linkage feature derived from the antecedent network's "\textbf{jumps}" output head in the cascade. \tabularnewline

\centering next stage (SGH) & \centering 1 & \centering C & Targeted activity transition ID derived from the antecedent network's "\textbf{next stage}" output head. \tabularnewline

\bottomrule
\end{tabular}
\end{table}

\subsection{Neural Network Input and Intermediate Representation}

As illustrated in the NNFC architecture (Figure~\ref{fig:feedback_with_inner_coupling}), each neural agent comprises three cascaded neural networks labeled A, B, and C. To ensure universal applicability across heterogeneous engineering domains, COOL processes intermediate symbolic structures at a standardized \textbf{Intermediate Representation (IR)} level using \textbf{Three-Address Code (TAC)}. This abstraction is critical as it decouples the neuro-symbolic reasoning process from the syntactic constraints of specific DSLs or the low-level machine code of the execution platform~\cite{sujeeth2014delite}. The TAC-based representation is subsequently transformed into a graph structure, preserving the topological semantics of the derivation state for graph-based neural processing.

The cascade of networks A, B, and C is a deliberate reliability-centric design. In this cascade, Network B ingests the symbolic state along with the SGH-head outputs from Network A; similarly, Network C processes outputs from Network B. This architecture facilitates \textbf{cross-network verification}: each downstream network is forced to evaluate the consistency of the upstream predictions. By enabling the accumulation of residual errors through the serial data flow, the system effectively amplifies the discrepancy of mispredictions, making logic-level inconsistencies easier to detect and filter by the cascade reliability layer.

The specific input features utilized by these networks are detailed in Table~\ref{tab:input_features}, providing a comprehensive encoding of the symbolic state and the regulatory context.

\begin{table}[!h]
\centering
\caption{Specification of neural network output features. These heads provide multi-objective optimizations, ranging from contextual task identification (TDH) and reachability assessment (SSPH) to fine-grained priority compensation (SGH).}
\label{tab:output_features}
\setlength{\tabcolsep}{3pt}
\renewcommand{\arraystretch}{1.5}

\begin{tabular}{p{0.17\textwidth}p{0.2\textwidth}p{0.10\textwidth}p{0.47\textwidth}}
\toprule
\centering \textbf{Output Head} & \centering \textbf{Dim.} & \centering \textbf{Network} & \centering \textbf{Signification} \tabularnewline
\midrule
\centering domain (TDH) & \centering 2 & \centering A, B, C & Contextual relevance of the current symbolic state to the active domain logic. \tabularnewline

\centering feasibility (SSPH) & \centering 2 & \centering A, B, C & Probabilistic assessment of the reachability of a complete derivation from the current state. \tabularnewline

\centering jumps (SGH) & \centering max\_tree\_depth *3 & \centering A, B, C & Structural path descriptors defining the precise target node within the syntax tree for rule application. \tabularnewline

\centering next stage (SGH) & \centering 1 & \centering A, B, C & The targeted sub-DSL activity ID for the next step in the CoL coordination chain. \tabularnewline

\centering heuristic sign (SGH) & \centering 2 & \centering A, B, C & Polarized direction (positive/negative) of the neural compensation for the expert heuristic. \tabularnewline

\centering heuristic value (SGH) & \centering 1 & \centering A, B, C & Numerical magnitude of the dynamic compensation gain for the baseline rule priority. \tabularnewline

\centering expression (SGH) & \centering 2 & \centering A, B, C & Structural type classification of the rule consequence (complex expression vs. terminal). \tabularnewline

\bottomrule
\end{tabular}
\end{table}

\subsection{Hybrid Graph-Sequential Neural Architecture}

\begin{figure}[htbp]
    \centering
    \includegraphics[width=1\linewidth]{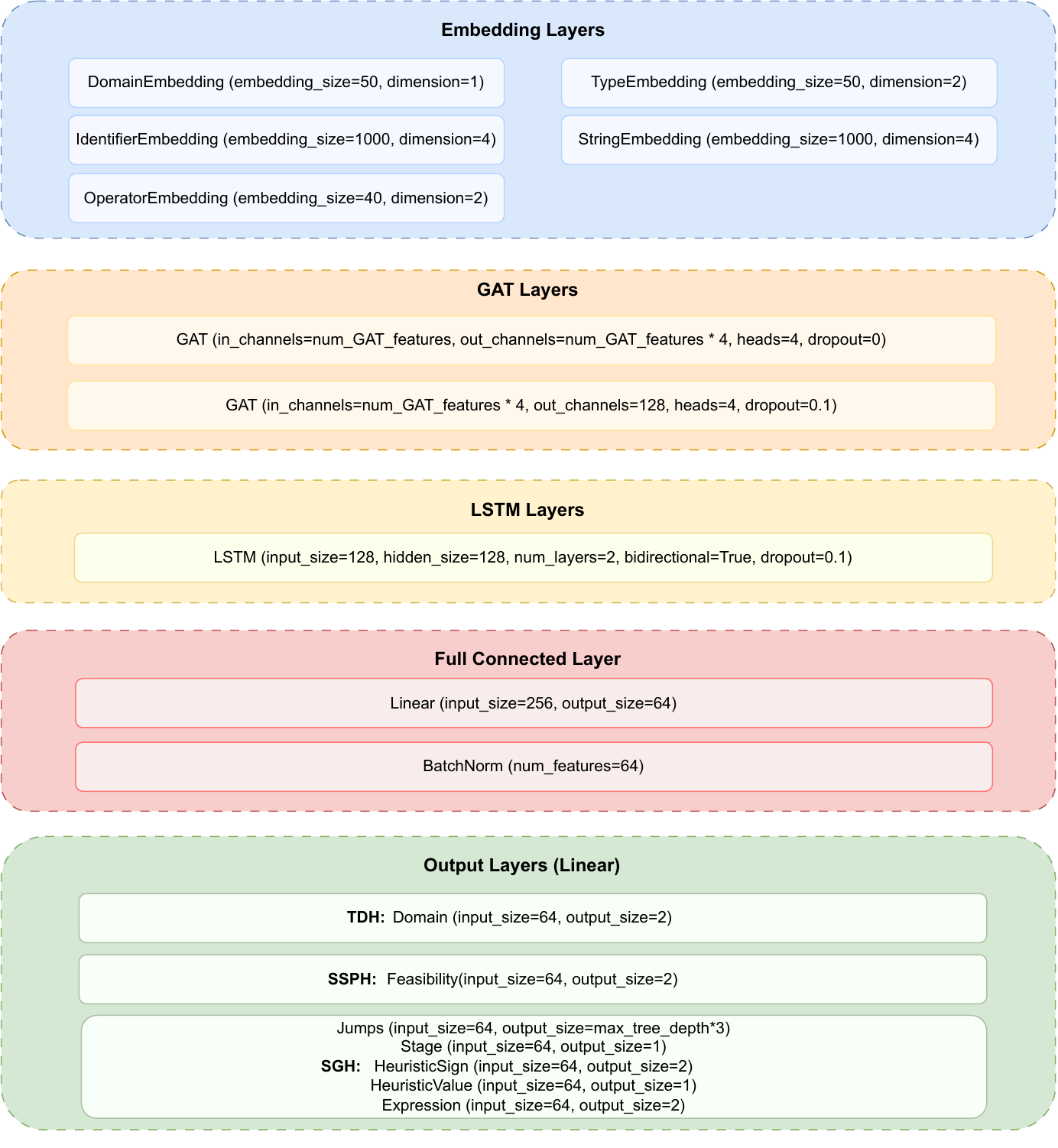}
    \caption{Detailed layer architecture of the COOL neural agent. The model facilitates a multi-stage processing pipeline: (1)~Semantic embedding of discrete symbolic tokens; (2)~Topological feature extraction via GAT layers to capture structural dependencies within the TAC graph; (3)~Contextual dependency modeling via Bi-LSTM to resolve sequential logic; and (4)~Decoupled inference through multi-head output layers for task detection (TDH), reachability pruning (SSPH), and strategy guidance (SGH).}
    \label{fig:neural_network_layers}
\end{figure}

Since the Three-Address Code (TAC) inherently captures both the topological dependencies (data-flow graphs) and the sequential directives (instruction streams) of a symbolic derivation, the neural agent is designed as a hybrid architecture capable of simultaneous spatial and temporal modeling. As illustrated in Figure~\ref{fig:neural_network_layers}, the architecture integrates Graph Attention Networks (GAT) with Bidirectional Long Short-Term Memory (Bi-LSTM) units to achieve high-fidelity feature extraction.

The end-to-end processing flow within each modular neural agent consists of the following functional stages:

\begin{enumerate}
    \item \textbf{High-Dimensional Semantic Embedding:} Categorical inputs (e.g., domains, identifiers, operators) are first mapped into dense, continuous vector spaces through learnable embedding layers. This stage suppresses noise and prepares the symbolic tokens for high-performance tensor operations~\cite{hrinchuk2019tensorized}.
    \item \textbf{Topological Feature Extraction via GAT:} The engine constructs a graph representation from the TAC lines, where Graph Attention (GAT) layers are employed to extract intricate structural details. By assigning varying attention weights to neighboring nodes, the agent focuses on critical data-flow interactions that dictate the success of a derivation rule~\cite{velickovic2017graph,wu2022graph}.
    \item \textbf{Contextual Modeling via Bidirectional LSTM:} Recognizing the sequential nature of derivation trajectories, we adopt a Bidirectional LSTM (Bi-LSTM) layer following the graph extraction. This dual-directional approach allows the agent to enrich its contextual understanding by simultaneously considering the historical trajectory and the potential look-ahead features of the TAC sequence~\cite{huang2015bidirectional,nye2020learning}.
    \item \textbf{Decoupled Multi-Task Inference:} The aggregated spatial-temporal features are channeled through independent output heads. This decoupling ensures that specialized tasks—such as contextual scoping (TDH) and feasibility assessment (SSPH)—do not interfere with the fine-grained heuristic compensation logic (SGH).
\end{enumerate}

The internal reliability of the NNFC mechanism is further reinforced by the cascade (series coupling) of three identical neural units: A, B, and C. In this sequential arrangement, the downstream networks inherit key strategy descriptors from their predecessors. Specifically, Network B incorporates the "\textbf{jumps}" descriptor from Network A, while Network C integrates both "\textbf{jumps}" and "\textbf{next stage}" features from Network B. This recursive inheritance enables the cascade to amplify representation discrepancies in erroneous paths, providing the deterministic filtering required for high-stakes engineering regulation. The consistency of output features across all cascaded units (as specified in Table~\ref{tab:output_features}) ensures that the reliability layer can perform effective cross-network consensus verification.

\subsection{Prediction Filtering and Discrepancy Detection}
\label{app:filter}

As established in the NNFC cascade architecture, the reliability of neural guidance is verified by evaluating the output consistency across the sequentially coupled neural units A, B, and C. This mechanism identifies potential mispredictions by monitoring the discrepancy of the multi-head outputs. For discrete classification tasks (e.g., TDH and stage prediction), the filter enforces a \textit{consensus requirement}: the prediction is only accepted if all cascaded units generate identical outputs. For continuous regression tasks (e.g., heuristic value compensation), a \textit{tolerance-based filtering} approach is adopted, where the prediction is discarded if the relative discrepancy between units exceeds a 10\% threshold. This process ensures that only high-confidence neural signals are allowed to influence the symbolic reasoning flow.

\subsection{Neural Guidance Acting on Heuristic Values}

The expert-defined baseline heuristics ($\mathbf{h}$) within each sub-DSL represent the structural knowledge of domain experts. The neural agent provides dynamic compensation to these baselines, refining the rule application priorities in real-time based on the outputs of its functional heads:

\begin{itemize}
    \item \textbf{Task Detection Head (TDH):} If TDH determines that the current symbolic state deviates from the active CoL activity's processing scope, the system applies a \textit{contextual penalty} to all rules within that state to suppress unauthorized rule applications: $\mathbf{h}[i] = \mathbf{h}[i] - |\mathbf{h}[i]| - 10$.
    
    \item \textbf{Search Space Prune Head (SSPH):} For states identified as within-domain by TDH, the SSPH assesses their feasibility. If a state is predicted to be unreachable from the target solution, a \textit{pruning penalty} is applied to all associated rule priorities: $\mathbf{h}[i] = \mathbf{h}[i] - |\mathbf{h}[i]| - 10$.
    
    \item \textbf{Search Guidance Head (SGH):} For feasible derivation paths, SGH provides fine-grained priority adjustments. If the current rule's features align with the neural strategy (i.e., matching logical values and numerical parameters within a $\pm 10\%$ range), it receives a \textit{positive compensation bonus}: $\mathbf{h}[i] = \mathbf{h}[i] + |\mathbf{h}[i]|$. Otherwise, the rule priority is suppressed via a \textit{guidance penalty}: $\mathbf{h}[i] = \mathbf{h}[i] - |\mathbf{h}[i]| - 10$.
\end{itemize}

\section{Signal Clipper for Multi-instance Coordination}
\label{app:clipper}

The Signal Clipper serves as a consistency guard for multi-DSL regulation, particularly when managing coordination between multiple CoL instances. It modulates the expert-defined signals by capping those that conflict with the self-corrected neural guidance to zero, thereby refining the active search envelope:

\begin{equation}
u_2 = \begin{cases}
0 & \text{if } u_1 > 0 \text{ and the current CoL rule doesn't align with} \\
 &\text{ the agent guidance, while there exists another CoL instance} \\
& \text{ whose rule aligns with the guidance within the search space} \\
u_1 & \text{otherwise}
\end{cases}
\end{equation}

By suppressing inconsistent rule-application signals ($u_2 \to 0$), the Clipper effectively regulates the interactions between heterogeneous DSL modules, preventing unnecessary interleaving and ensuring that the global reasoning process remains focused on the most efficient derivation trajectories.

\section{Experimental Benchmark Specifications}
\label{app:experiment}

To evaluate the neuro-symbolic regulation capabilities of the COOL framework, we selected benchmarks that represent rule-intensive engineering logic. These specifications are implemented in the COOL host language to verify the framework's efficiency in managing modular DSLs. The \textit{relational} tasks are adapted from the CLUTRR~\cite{sinha2019clutrr} dataset, focusing on formal logical inference, while the \textit{symbolic} tasks involve algebraic tasks generated by GPT~\cite{achiam2023gpt}. For the relational benchmarks, we utilize the formal representation provided by CLUTRR to ensure deterministic evaluation.

\subsection{Benchmark Task Examples}

The following examples illustrate the different coordination scenarios used in our study:

\subsubsection*{Relational Reasoning Specifications}

These tasks evaluate the coordination efficiency of sub-DSLs within a single task domain, where the logical search space allows for \textit{completeness guarantees}:

\asumr{Relational Task Example}{}{
\hspace{-5pt}\textcolor{gray}{//load 4 DSLs for family relationship reasoning} \\
\textcolor{teal}{\#load(family) }\\
\\
\textcolor{gray}{//Relational reasoning questions like (50 per batch):}\\
(Wesley) is (James)s son \textcolor{blue}{\&} (Martha) is (Wesley)s daughter \textcolor{blue}{\&} (Hugh) is (Martha)s uncle \textcolor{blue}{\&} (Hugh) is (James)s (\textcolor{olive}{\${}relation});\\
...
}

\subsubsection*{Symbolic Transformation Specifications}

These tasks assess the framework's ability to regulate sub-DSLs in high-depth, algebraic synthesis environments where \textit{completeness is not guaranteed} due to the complexity of the transformation rules:

\asumr{Symbolic Task Example}{}{
\hspace{-5pt}\textcolor{gray}{//load 7 DSLs for family relationship reasoning and symbolic reasoning}\\
\textcolor{teal}{\#load(quadratic) }\\
\\
\textcolor{gray}{//Symbolic reasoning questions like (50 per batch):}\\
\textcolor{olive}{\$x}\^{}2 + 4*\textcolor{olive}{\$x} == 3;\\
...
}

\subsubsection*{Heterogeneous Cross-Domain Specifications}

Used to evaluate the collaboration efficiency across distinct task domains, representing complex engineering scenarios where relational inference and algebraic computation must be interleaved:

\asumr{Cross-Type Task Example}{}{
\hspace{-5pt}\textcolor{gray}{//load 11 DSLs for relational reasoning and symbolic reasoning}\\
\textcolor{teal}{\#load(family)}\\
\textcolor{teal}{\#load(quadratic)}\\ 
\\
\textcolor{gray}{//Symbolic reasoning questions like (50 per batch):}\\
\textcolor{olive}{\$x}\^{}2 + 4*\textcolor{olive}{\$x} == 3;\\
...\\
\textcolor{gray}{//Relational reasoning questions like (50 per batch):}
(Wesley) is (James)s son \textcolor{blue}{\&} (Martha) is (Wesley)s daughter \textcolor{blue}{\&} (Hugh) is (Martha)s uncle \textcolor{blue}{\&} (Hugh) is (James)s (\textcolor{olive}{\$relation});\\
...
}

In our experimental methodology, execution logic that does not directly contribute to the task specifications is treated as a control variable and is systematically deducted from the final performance results to ensure an unbiased assessment of the multi-DSL regulation overhead.

\subsection{Structural Configuration of Modular DSLs}

The granularity of the divide-and-conquer strategy in COOL is reflected in the structural configuration of its constituent sub-DSLs. Table~\ref{tab:characteristics_of_kdls} summarizes the distribution of the rule-based operational primitives utilized in our benchmarks. These rules are categorized according to their functional roles—such as production, reduction, recursion, and permutation—representing the diverse logical challenges encountered in rule-intensive engineering regulation.

\begin{table}[H]
\centering
\caption{Structural complexity and rule distribution of the multi-DSL configurations. Each benchmark is decomposed into a specific number of sub-DSLs (Activities), each managing a subset of domain-specific logic primitives.}

\label{tab:characteristics_of_kdls}
\setlength{\tabcolsep}{4pt}

\begin{tabular}{m{0.16\textwidth}|m{0.05\textwidth}m{0.14\textwidth}m{0.13\textwidth}m{0.12\textwidth}m{0.175\textwidth}|m{0.11\textwidth}}
\toprule 
\multirow{2}{*}{\centering \textbf{Benchmark}} & \multicolumn{5}{c|}{\centering \textbf{Rule-set Distribution}} & \multirow{2}{*}{\centering \textbf{sub-DSLs}} \tabularnewline
 & \centering \textbf{Total} & \centering \textbf{Production Rules} & \centering \textbf{Reduction Rules} & \centering \textbf{Recursive Rules} & \centering \textbf{Permutation Rules} & \tabularnewline
\midrule 

\centering Relational & 
\centering 40 & 
\centering 36 & 
\centering 2 & 
\centering 16 & 
\centering 0 & 
\centering 4
\tabularnewline

\centering Symbolic & 
\centering 55 & 
\centering 17 & 
\centering 26 & 
\centering 3 & 
\centering 11 & 
\centering 7\tabularnewline

\centering Cross-task & 
\centering 95 & 
\centering 53 & 
\centering 28 & 
\centering 19 & 
\centering 11 & 
\centering 11\tabularnewline
\bottomrule
\end{tabular}

\end{table}

The configuration shows that while relational tasks primarily involve production and recursive logic for identity tracing, symbolic tasks require a higher proportion of reduction and permutation rules for algebraic simplification. The cross-task benchmark represents a full-scale integration of these heterogeneous logic domains, providing a rigorous testbed for the CoL orchestration and NNFC adaptive compensation mechanisms.

\section{System Implementation and Engineering Realization}

\subsection{Execution Engine and Toolchain Architecture}

To bridge the gap between high-level regulatory logic and execution-level efficiency, the COOL framework is developed as an autonomous, high-performance execution engine. Unlike generic DSL workbenches, COOL is implemented from the ground up in \textbf{C++} to ensure the low-latency handling of intensive tree manipulation and symbolic state transformations. For syntax orchestration and semantic parsing across heterogeneous DSLs, we employ \textbf{Lex}~\cite{lesk1975lex} and \textbf{YACC}~\cite{johnson1975yacc}, providing a deterministic foundation for rule-intensive applications. The adaptive compensation layer (NNFC) is developed in \textbf{Python} using the \textbf{PyTorch} ecosystem, facilitating the modular deployment of distributed neural agents. This hybrid toolchain provides a seamless integration between deterministic symbolic logic and adaptive AI feedback. 

The significant engineering effort invested in the framework is quantified in Table~\ref{tab:code_effort}. The primary C++ codebase forms the high-speed execution substrate, while the Lex/YACC components handle the precise orchestration of domain-specific grammars. (GitHub repository at \url{https://coolang.org})

\begin{table}[h]
    \centering
     \caption{Engineering development metrics of the COOL framework. The implementation emphasizes architectural autonomy and low-latency execution for complex regulation tasks.}
     \label{tab:code_effort}
    \begin{tabular}{>{\centering\arraybackslash}p{0.23\textwidth}>{\centering\arraybackslash}p{0.23\textwidth}>{\centering\arraybackslash}p{0.33\textwidth}}
        \toprule
        \textbf{Language} & \textbf{SLOC} & \textbf{Core Components} \\
        \midrule
        C++ & 60,000 & High-performance execution engine \\
        Python & 3,000 & Distributed neural agents (NNFC) \\
        Lex & 1,000 & Deterministic syntactic parsers \\
        YACC & 2,000 & Hierarchical semantic parsers \\
        \bottomrule
    \end{tabular}
\end{table}

\subsection{Source Code Implementation in COOL}

To further elucidate the operational mechanics of the COOL framework, this section provides representative implementation paradigms within the COOL host language. 

The following source code subsections showcase three critical layers of the engineering realization: 
(1)~\textbf{Symbolic Rule Definitions and Heuristic Encoding}, which demonstrate how domain expertise is structured via heuristic vectors; 
(2)~\textbf{Benchmark Specification Programs}, representing the high-level input logic used in our relational and symbolic experiments; and 
(3)~\textbf{TAC-based Intermediate Representation}, which details the execution-level instructions that facilitate cross-DSL coordination and neural feature extraction.

\subsubsection{Relational Task DSLs}
\label{app:relational_implementation}

\begin{Verbatim}[breaklines=true]
//1 Separate Relations and Genders
expr:@(9){(a) is (b)s grandson}{
    return:(a) is male &  (a) is (b)s grandchild & (b) is (a)s grandparent;
}
...

//2 Reason Inverse Relations
expr:@(0,7,3){(a) is (b)s grandchild}{
    if(this expr.exist subexpr{(b) is (a)s grandparent} == false){
        return: (a) is (b)s grandchild & (b) is (a)s grandparent;
    }
    abort;
}
...

//3 Reason Indirect Relations
expr:@(0,0,5){(a) is (b)s sibling}{
    placeholder:p1;
    while(this expr.find subexpr{(p1) is (a)s sibling}){
        if(this expr.exist subexpr{(p1) is (b)s sibling} == false && p1 != b){
return: (a) is (b)s sibling & (p1) is (b)s sibling;
        }
        p1.reset();
    }
    p1.reset();
    while(this expr.find subexpr{(p1) is (a)s parent}){
        if(this expr.exist subexpr{(p1) is (b)s parent} == false){
return: (a) is (b)s sibling & (p1) is (b)s parent;
        }
        p1.reset();
    }
    p1.reset();
    while(this expr.find subexpr{(p1) is (a)s pibling}){
        if(this expr.exist subexpr{(p1) is (b)s pibling} == false){
return: (a) is (b)s sibling & (p1) is (b)s pibling;
        }
        p1.reset();
    }
    p1.reset();
    while(this expr.find subexpr{(p1) is (a)s grandparent}){
        if(this expr.exist subexpr{(p1) is (b)s grandparent} == false){
return: (a) is (b)s sibling & (p1) is (b)s grandparent;
        }
        p1.reset();
    }
    p1.reset();
    abort;
}
...

//4 Recombine Relations and Genders, Eliminate Irrelevant Relations
expr:@(0,0,0,8){(a) is (b)s ($relation)}{
    //immediate family
    placeholder:p1;
    while(this expr.find subexpr{(a) is (b)s grandchild}){    
        if(this expr. exist subexpr{(a) is male}){
return: $relation == "grandson";
        }
        if(this expr.exist subexpr{(a) is female}){
return:$relation == "granddaughter";
        }
        p1.reset();
    }
    p1.reset();
    while(this expr.find subexpr{(a) is (b)s child}){    
        if(this expr. exist subexpr{(a) is male}){
return: $relation == "son";
        }
        if(this expr.exist subexpr{(a) is female}){
return:$relation == "daughter";
        }
        p1.reset();
    }
    ...
    abort;
}
...
expr:@(0,0,0,10){a & ($b == c)}{
    return:b == c;
}
...
\end{Verbatim}

\subsubsection{Symbolic Task DSLs}
\label{app:symbolic_dsl}
\begin{Verbatim}[breaklines=true]
// Common Transformations
expr:@(2,2,2,2,2){0+#a}{
    return:a;
}
expr:@(2,2,2,2,2){#a+0}{
    return:a;
}
...

// 1 Expand Square Terms
expr:@(5,0,0,0){(#?a + #?b)^2}{
    return:a^2+2*a*b+b^2;
}
expr:@(5,0,0,0){(#?a - #?b)^2}{
    return:a^2+(-2)*a*b+b^2;
}
expr:@(6,0,0,0){(#a*#b)^2}{
    return:a^2*b^2;
}
...

// 2 Expand Bracketed Terms
expr:@(0,4,0,0,0){#?a-(#?b+#?c)}{
    return:a-b-c;
}
expr:@(0,3.8,0,0,0){(#?b+#?c)*#?a}{
    return:b*a+c*a;
}
...

// 3 Extract Coefficients
expr:@(0,0,5,0){$x*a}{
    return:a*x;
}
expr:@(0,0,4.8,0){(immediate:a*$x)*(immediate:b*$x)}{
    new:tmp = a*b;
    return:tmp*x^2;
}
expr:@(0,0,4.6,0){$x*(a*$x)}{
    return:a*x^2;
}
...

// 4 Re-Express Negative Coefficients
expr:@(0,0,0,3.5,0){#a-$x}{
    placeholder:p1;
    placeholder:p2;
    if(x.exist subexpr{p1*p2}){
        abort;
    }
    return:a+(-1)*x;
}
expr:@(0,0,0,3.7,0){#a-immediate:b*$x}{
    new:tmp = 0 - b;
    return:a+tmp*x;
}
...

//5 Arrange Terms in Descending Order, Combine Like Terms
expr:@(0,0,0,0,3){immediate:a*$x+immediate:b*$x}{
    new:tmp = a+b;
    return:tmp*x;
}
expr:@(0,0,0,0,2.8){a1*$x+a2*$x^2}{
    return:a2*x^2+a1*x;
}
...

//6 Convert to Standard Form
expr:@(0,0,0,0,0,2.5){a*$x^2+b*x == #d}{
    return: a*$x^2+b*x + 0 == d;
    
}
expr:@(0,0,0,0,0,2.5){b*$x == $d}{
    
    if(d.exist subexpr{x^2}){
        return: 0*x^2 + b*x + 0 == d;
    }else {
        abort;
    }
}
expr:@(0,0,0,0,0,-4){$a==$b}{
    return:b==a;
}
...

//7 Apply Solution Formula
@(0,0,0,0,0,0,0,10){a*$x^2+b*x+c==0}{
    if(b^2-4*a*c<0){
        x="null";
    }
    else {
        new:x1=(-b+(b^2-4*a*c)^0.5)/(2*a);
        new:x2=(-b-(b^2-4*a*c)^0.5)/(2*a);
        x={x1,x2};    
    }
};
\end{Verbatim}

\subsubsection{Relational Tasks at Difficulty Level A}
\begin{Verbatim}[breaklines=true]
#load(family) // Load the CoL DSL library for Relational Tasks
new:relation = "";
// [Francisco]'s brother, [Wesley], recently got elected as a senator. [Lena] was unhappy with her son, [Charles], and his grades. She enlisted a tutor to help him. [Wesley] decided to give his son [Charles], for his birthday, the latest version of Apple watch. 
// Ans: (Francisco) is (Lena)s brother 
new:Lena = "Lena"; 
new:Charles = "Charles"; 
new:Wesley = "Wesley"; 
new:Francisco = "Francisco";
(Charles) is (Lena)s son & (Wesley) is (Charles)s father & (Francisco) is (Wesley)s brother & (Francisco) is (Lena)s ($relation); 
relation-->"#FILE(SCREEN)";

// [Clarence] woke up and said hello to his wife, [Juanita]. [Lynn] went shopping with her daughter [Felicia]. [Felicia]'s sister [Juanita] was too busy to join them. 
// Ans: (Lynn) is (Clarence)s mother-in-law 
new:Clarence = "Clarence";
new:Juanita = "Juanita";
new:Felicia = "Felicia";
new:Lynn = "Lynn";
(Juanita) is (Clarence)s wife & (Felicia) is (Juanita)s sister & (Lynn) is (Felicia)s mother & (Lynn) is (Clarence)s ($relation); 
relation-->"#FILE(SCREEN)";
...
\end{Verbatim}

\subsubsection{Relational Tasks at Difficulty Level B}
\begin{Verbatim}[breaklines=true]
#load(family) // Load the CoL DSL library for Relational Tasks
new:relation = "";
// [Antonio] was happy that his son [Bernardo] was doing well in college. [Dorothy] is a woman with a sister named [Tracy]. [Dorothy] and her son [Roberto] went to the zoo and then out to dinner yesterday. [Tracy] and her son [Bernardo] had lunch together at a local Chinese restaurant. 
// Ans: (Roberto) is (Antonio)s nephew 
new:Antonio = "Antonio";
new:Bernardo = "Bernardo"; 
new:Tracy = "Tracy"; 
new:Dorothy = "Dorothy"; 
new:Roberto = "Roberto";
(Bernardo) is (Antonio)s son & (Tracy) is (Bernardo)s mother & (Dorothy) is (Tracy)s sister & (Roberto) is (Dorothy)s son & (Roberto) is (Antonio)s ($relation); 
relation-->"#FILE(SCREEN)";

// [Bernardo] and his brother [Bobby] were rough-housing. [Tracy], [Bobby]'s mother, called from the other room and told them to play nice. [Aaron] took his brother [Bernardo] out to get drinks after a long work week. [Tracy] has a son called [Bobby]. Each day they go to the park after school. ans: (Bobby) is (Aaron)s brother 
new:Aaron = "Aaron"; 
new:Bernardo = "Bernardo"; 
new:Bobby = "Bobby"; 
new:Tracy = "Tracy";
(Bernardo) is (Aaron)s brother & (Bobby) is (Bernardo)s brother & (Tracy) is (Bobby)s mother & (Bobby) is (Tracy)s son & (Bobby) is (Aaron)s ($relation); 
relation-->"#FILE(SCREEN)";
...
\end{Verbatim}

\subsubsection{Symbolic Tasks at Difficulty Level A}
\begin{Verbatim}[breaklines=true]
#load(quadratic) // Load the CoL DSL library for Symbolic Tasks
new:x = 1;
6*$x^2 == 3*x - 7;
x-->"#FILE(SCREEN)";
($x - 6)*(x + 3) == x;
x-->"#FILE(SCREEN)";
...
\end{Verbatim}

\subsubsection{Symbolic Tasks at Difficulty Level B}
\begin{Verbatim}[breaklines=true]
#load(quadratic) // Load the CoL DSL library for Symbolic Tasks
new:x = 1;
$x*($x + 11) == 16*($x + 22); 
x-->"#FILE(SCREEN)";
$x*(36*$x + 50) - 11*(19 - 30*$x) == $x^2; 
x-->"#FILE(SCREEN)";
...
\end{Verbatim}

\subsubsection{Cross-Type Tasks}
\begin{Verbatim}[breaklines=true]
#load(quadratic) // Load the CoL DSL library for Symbolic Tasks
#load(family) // Load the CoL DSL library for Relational Tasks
new:x = 1;
$x^2 - 4*$x == 6; 
x --> "#FILE(SCREEN)";
...
new:relation = "";
// [Dolores] and her husband [Don] went on a trip to the Netherlands last year. [Joshua] has been a lovely father of [Don] and has a wife named [Lynn] who is always there for him. 
// Ans: (Dolores) is (Lynn)s daughter-in-law 
new:Lynn = "Lynn"; 
new:Joshua = "Joshua"; 
new:Don = "Don"; 
new:Dolores = "Dolores";
(Joshua) is (Lynn)s husband & (Don) is (Joshua)s son & (Dolores) is (Don)s wife & (Dolores) is (Lynn)s ($relation); 
relation-->"#FILE(SCREEN)";
...
\end{Verbatim}

\subsubsection{COOL Intermediate Representation}
\label{app:cool_ir}
\begin{Verbatim}[breaklines=true]
"codeTable": [
    {
        "boundtfdomain": "",
        "grounded": false,
        "operand1": {
            "argName": "x",
            "argType": "identifier",
            "changeable": 1,
            "className": "",
            "isClass": 0
        },
        "operand2": {
            "argName": "2",
            "argType": "number",
            "changeable": 0,
            "className": "",
            "isClass": 0
        },
        "operator": {
            "argName": "^",
            "argType": "other"
        },
        "result": {
            "argName": "1418.4",
            "argType": "identifier",
            "changeable": 1,
            "className": "",
            "isClass": 0
        },
        "root": false
    },
    ...
]
\end{Verbatim}

\end{document}